\documentclass{aa}  

\usepackage{graphicx}
\usepackage{hyperref}                                                           
\usepackage{natbib}
\usepackage[varg]{txfonts}

\makeatletter
\renewcommand*\aa@pageof{, page \thepage{} of \pageref*{LastPage}}
\makeatother

\begin{document}

\title{An automated procedure for the detection of the Yarkovsky effect and results from the ESA NEO Coordination Centre\thanks{Full tables \ref{tab:det_short}, \ref{tab:chivalues} and \ref{tab:jpl_comparison} are only available in electronic form at the CDS via anonymous ftp to \url{cdsarc.u-strasbg.fr} (130.79.128.5) or via \url{http://cdsweb.u-strasbg.fr/cgi-bin/qcat?J/A+A/}.}}
\titlerunning{Automated detection of the Yarkovsky effect}

   \author{Marco Fenucci\inst{1,2}
          \and
          Marco Micheli \inst{1,3}
          \and
          Francesco Gianotto\inst{1,3}
          \and
          Laura Faggioli\inst{1,3}
          \and
          Dario Oliviero\inst{1,2}
          \and 
          Andrea Porru\inst{1,4}
          \and
          Regina Rudawska\inst{5,6}
          \and
          Juan Luis Cano\inst{1}
          \and
          Luca Conversi\inst{1}
          \and
          Richard Moissl\inst{1}
          }
   \authorrunning{M. Fenucci et al.}
   \institute{ESA NEO Coordination Centre, Largo Galileo Galilei, 1, 00044 Frascati, Italy\\
               \email{marco.fenucci@ext.esa.int}
    }

    \institute{ ESA ESRIN / PDO / NEO Coordination Centre, Largo Galileo Galilei, 1, 00044 Frascati (RM), Italy \\ \email{marco.fenucci@ext.esa.int}
    \and
    Elecnor Deimos, Via Giuseppe Verdi, 6, 28060 San Pietro Mosezzo (NO), Italy
    \and
    RHEA Systems, Via di Grotte Portella, 28, 00044 Frascati (RM), Italy
    \and
    Alia Space System Srl, Via San Giuseppe Calasanzio 15, 00044 Frascati (RM), Italy
    \and
    ESA ESTEC / PDO, Keplerlaan 1, 2201 AZ Noordwijk, The Netherlands
    \and
    RHEA Group, Jonckerweg 18, 2201 DZ Noordwijk, The Netherlands
    }

   \date{Received --- / Accepted ---}


  \abstract
  {The measurement of the Yarkovsky effect on near-Earth asteroids (NEAs) is
  common practice in orbit determination today, and the number of detections
  will increase with the developments of new and more accurate telescopic
  surveys. However, the process of finding new detections and identifying
  spurious ones is not yet automated, and it often relies on personal judgment.
  }
  {We aim to introduce a more automated procedure that can search for NEA
  candidates to measure the Yarkovsky effect, and that can identify spurious
  detections. }
  {The expected semi-major axis drift on an NEA caused by the Yarkovsky effect
  was computed with a Monte Carlo method on a statistical model of the physical
  parameters of the asteroid that relies on the most recent NEA population
  models and data. The expected drift was used to select candidates in which
  the Yarkovsky effect might be detected, according to the current knowledge of
  their orbit and the length of their observational arc. Then, a
  nongravitational acceleration along the transverse direction was estimated
  through orbit determination for each candidate. If the detected
  acceleration was statistically significant, we performed a statistical test
  to determine whether it was compatible with the Yarkovsky effect model.
  Finally, we determined the dependence on an isolated tracklet. }
  {Among the known NEAs, our procedure automatically found 348 detections of
  the Yarkovsky effect that were accepted. The results are overall compatible
  with the predicted trend with the the inverse of the diameter, and the
  procedure appears to be efficient in identifying and rejecting spurious
  detections. This algorithm is now adopted by the ESA NEO Coordination Centre
  to periodically update the catalogue of NEAs with a measurable Yarkovsky
  effect, and the results are automatically posted on the web portal.}
  {}

     \keywords{Astrometry and celestial mechanics - Minor planets, asteroids: general - Methods: statistical}


   \maketitle

\section{Introduction}
\label{s:intro}
The Yarkovsky effect is a nongravitational force caused by thermal emission. It affects the dynamics of asteroids whose diameter is smaller than about 40 km \citep[see e.g.][]{vokrouhlicky_1998, bottke-etal_2006, vokrouhlicky-etal_2015}, and it mainly manifests as a drift in the semi-major axis of the object. It is known to cause several phenomena in the dynamics of small bodies in the Solar System, such as the spreading of asteroid families \citep[see e.g.][]{spoto-etal_2015}, and the transport of objects from the main belt to the near-Earth region \citep[see e.g.][]{granvik-etal_2017}. From the point of view of planetary defence, measuring nongravitational effects acting on near-Earth asteroids (NEAs) is important for a reliable assessment of the impact risk, especially for the long-time horizon search of potential impacts \citep{chesley-etal_2014, farnocchia-chesley_2014, spoto-etal_2014, vokrouhlicky-etal_2015b}. Additionally, measurements of the Yarkovsky effect can be used to extrapolate the physical properties of the asteroids \citep[see e.g.][]{chesley-etal_2014, rozitis-etal_2014, fenucci-etal_2021, fenucci-etal_2023}, which are key parameters to know for the design of lander and sample-return spacecraft missions \citep{murdoch-etal_2021}, or to predict the efficiency of a kinetic impactor-deflection mission \citep{cheng-etal_2023, daly-etal_2023, thomas-etal_2023}.   

Since the first measurement of the Yarkovsky effect on asteroid (6489) Golevka
\citep{chesley-etal_2003}, many detections on NEAs have been found and
announced in a series of related works \citep{nugent-etal_2012,
farnocchia-etal_2013, delvigna-etal_2018, greenberg-etal_2020}. Today, the
information about the detections is regularly provided by the three main
centers for asteroid orbit computation: the ESA NEO Coordination
Centre\footnote{\url{https://neo.ssa.esa.int/}} (NEOCC), the
NEODyS\footnote{\url{https://newton.spacedys.com/}} service, and the NASA JPL
Solar System Dynamics\footnote{\url{https://ssd.jpl.nasa.gov/}} (SSD) group.
However, the process of detecting new measurements is not fully automated, and
the identification of spurious detections still relies on manual checks that
are often based on personal judgment. There are several reasons that make a
detection spurious, such as dynamical modeling issues or poor-quality old
astrometry. Some detections may also appear to be incompatible with the
Yarkovsky effect, but they are in fact real and are caused by other poorly
understood phenomena, as in the case of asteroid (523599) 2003~RM
\citep{farnocchia-etal_2023}. Therefore, in some cases it is questionable whether a Yarkovsky effect has really been detected or not.
On the other hand, the new and improved observational technologies and the
high-quality data provided by the ESA Gaia mission \citep{tanga-etal_2022} mean
that the number of detected Yarkovsky effects is only expected to increase in
the future. This will become especially true after the beginning of the
operational phase of new-generation ground-based surveys, such as the ESA
Flyeye telescope \citep{conversi-etal_2021} and the Vera Rubin Observatory
\citep{ivezic-etal_2019}, and of the the space-based infrared telescopes NEO
Surveyor by NASA \citep{mainzer-etal_2021} and NEOMIR by ESA
\citep{conversi-etal_2023}. These new facilities are expected to increase the
number of known asteroids by roughly an order of magnitude
\citep{jones-etal_2018}. This situation calls for the development of new and
automated methods for the detection of the Yarkovsky effect and for the
identification of spurious measurements. 

In contrast to previous efforts, we propose a systematic procedure here to search for candidates for measurements of the Yarkovsky effect on NEAs, and for the identification of spurious detections. We use a statistical model of the physical parameters of the asteroids that relies on the most recent NEA data sets and population models \citep{granvik-etal_2018, morbidelli-etal_2020, berthier-2022}. We first estimate the expected Yarkovsky effect with a Monte Carlo method. Then, a list of NEA candidates for the detection of the Yarkovsky effect is determined according to the expected semi-major axis drift, the observational arc, and the uncertainty of the current orbit. For each object of the list, a nongravitational force along the transverse direction is estimated through orbit determination, and only detections with a good signal-to-noise ratio that are statistically compatible with the prediction from the physical model are kept. Then, we determine the dependence of the detections on isolated tracklets, and only those that passed the check are accepted. 
Among known NEAs, the algorithm identified 348 detections. The procedure described here is now adopted by the ESA NEO Coordination Centre, and data are automatically uploaded on the web portal. 

The paper is organized as follows. In Sec.~\ref{s:methods} we describe the model for the Yarkovsky effect and the model of the physical parameters of the NEAs that we used to predict the semi-major axis drift, the method we used to determine the orbit, and the procedure we used to automatically detect the Yarkovsky effect. In Sec.~\ref{s:res} we provide the results we obtained by running the procedure on the known NEA population, we compare them with the results reported by the JPL SSD group, and we discuss the effectiveness and limitations of our algorithm. Finally, we summarize our conclusions in Sec.~\ref{s:conclusions}.

\section{Methods}
\label{s:methods}

\subsection{Orbit determination}
\label{ss:OD}
The dynamical model we used to determine the orbit of a NEA includes the gravitational perturbations of the Sun, the eight planets, the Moon, the 16 most massive main-belt asteroids, and Pluto. The masses and positions of all these bodies were all retrieved from the JPL ephemeris DE441 \citep{park-etal_2021}. The masses of the 16 main-belt asteroids included in the dynamical model and the mass of Pluto are given in Table~\ref{tab:masses_small_bodies}. Relativistic effects of the Sun, the planets, and the Moon were also added to the dynamical model as a first-order Newtonian expansion \citep{will_1993}. 
The Yarkovsky effect is expressed as an acceleration along the transverse direction $\hat{\mathbf{t}}$ of the form
\begin{equation}
    \mathbf{a}_t = A_2 \bigg(\frac{1 \textrm{ au}}{r}\bigg)^2\hat{\mathbf{t}},
    \label{eq:yarko_od_model}
\end{equation}
where $r$ is the distance from the Sun \citep{farnocchia-etal_2013}. The corresponding semi-major axis drift is obtained through the Gauss planetary equations by
\begin{equation}
    \frac{\text{d}a}{\text{d}t} = \frac{2 A_2 (1-e^2)}{np^2},
    \label{eq:A22dadt}
\end{equation}
where $n$ is the mean motion, and $p=a(1-e^2)$ is the semilatus rectum. 
Additionally, it is possible to optionally add solar radiation pressure (SRP) to the dynamical model. The SRP is modeled as a radial force $\mathbf{a}_r$ of the form
\begin{equation}
    \mathbf{a}_r = A_1 \bigg(\frac{1 \textrm{ au}}{r}\bigg)^2 \hat{\mathbf{r}},
\end{equation}
where $\hat{\mathbf{r}}=\mathbf{r}/r$ is the radial unit vector, and $\mathbf{r}$ is the heliocentric position of the asteroid. The coefficient $A_1$ is computed as in \citet{montenbruck-gill_2000},
\begin{equation}
    A_1 = (1-\Theta)\times \frac{\phi}{c} \times \frac{A}{m},
\end{equation}
where $\phi = 1.361$ kW m$^{-2}$ is the solar radiation energy flux at 1 au, $c$ is the speed of light, $A/m$ is the area-to-mass parameter, and $\Theta$ is the occultation function, which determines whether the object is occulted by a major planet. 

The parameter $A_2$ (and possibly the $A/m$ parameter, when SRP is included in the model) is determined together with the orbital elements by fitting the model to the observations through a least-squares procedure \citep{milani-gronchi_2009}. Observational outliers were identified and discarded with an automatic rejection algorithm developed by \citet{carpino-etal_2003}. In all the computations, we used the weighting scheme based on observatory statistics developed by \citet{veres-etal_2017}, and the debiasing scheme based on star catalogs developed by \citet{farnocchia-etal_2015}. The orbital elements estimated from the fit refer to the weighted mean of the observation times, as indicated in \citet{milani-gronchi_2009}. To perform the orbital fit, we used the ESA Aegis orbit determination and impact monitoring software \citep{faggioli-etal_2023}, developed and maintained by SpaceDyS s.r.l.\footnote{\url{https://www.spacedys.com/}} under ESA contracts.

\begin{table}[!ht]
    \renewcommand{\arraystretch}{1.3}
        \caption{Gravitational parameters $GM$ of the 16 most massive main-belt asteroids included in the dynamical model, and that of Pluto.}
    \centering
    \begin{tabular}{ll}
    \hline
    \hline
    Asteroid & $GM/GM_\odot$ \\
    \hline
 (1) Ceres          & 4.7191422767$\times 10^{-10}$  \\
 (2) Pallas         & 1.0297360324$\times 10^{-10}$  \\
 (3) Juno           & 1.4471670475$\times 10^{-11}$  \\
 (4) Vesta          & 1.3026836726$\times 10^{-10}$  \\
 (7) Hebe           & 8.5890389994$\times 10^{-12}$  \\
 (10) Hygea         & 4.2385986149$\times 10^{-11}$  \\
 (15) Eunomia       & 1.5243642468$\times 10^{-11}$  \\
 (16) Psyche        & 1.1978215785$\times 10^{-11}$  \\
 (31) Euphrosyne    & 8.1331596146$\times 10^{-12}$  \\
 (52) Europa        & 2.0216913527$\times 10^{-11}$  \\
 (65) Cybele        & 7.0687100328$\times 10^{-12}$  \\
 (87) Sylvia        & 1.6337820878$\times 10^{-11}$  \\
 (88) Thisbe        & 8.9653065564$\times 10^{-12}$  \\
 (107) Camilla      & 1.0878696848$\times 10^{-11}$  \\
 (511) Davida       & 2.9345275748$\times 10^{-11}$  \\
 (704) Interamnia   & 2.1327387534$\times 10^{-11}$  \\
 (134340) Pluto     & 7.3504789732$\times 10^{-9}$  \\
    \hline
    \end{tabular}
             \tablefoot{The gravitational parameters of the 16 main-belt asteroids are extracted from the header of the JPL ephemerides DE441 (\url{https://ssd.jpl.nasa.gov/ftp/eph/planets/ascii/de441/}). The gravitational parameter of Pluto is taken from \citet{brozovic-etal_2015}. The values are reported in the units of the gravitational parameter of the Sun, corresponding to $GM_\odot = 1.32712440040944 \times 10^{11}$ km$^3$ s$^{-2}$.}
    \label{tab:masses_small_bodies}
\end{table}

\subsection{Expected Yarkovsky drift}
\label{ss:expected_drift}

\subsubsection{Semi-analytical formula for the semi-major axis drift}
\label{sss:say}
The instantaneous semi-major axis drift produced by the Yarkovsky effect can be expressed by analytical formulas \citep{vokrouhlicky-etal_2017}, assuming a spherical body and a linearization of the boundary conditions.
It is given by
\begin{equation}
    \frac{\text{d}a}{\text{d}t} = \frac{2}{n^2 a} \mathbf{f}_{\text{Y}} \cdot \mathbf{v},
    \label{eq:dadt_i}
\end{equation}
where $a$ is the semi-major axis of the asteroid orbit, $\mathbf{v}$ is the heliocentric orbital velocity, and $\mathbf{f}_{\text{Y}}$ is the instantaneous value of the Yarkovsky acceleration.
The term $\mathbf{f}_{\text{Y}}$ is the result of the sum of the diurnal component $\mathbf{f}_{\text{Y,d}}$ and the seasonal component $\mathbf{f}_{\text{Y,s}}$.
The diurnal component is expressed as
\begin{equation}
    \mathbf{f}_{\text{Y, d}} = \kappa [(\hat{\mathbf{r}} \cdot \hat{\mathbf{s}})\hat{\mathbf{s}} + \gamma_1 (\hat{\mathbf{r}} \times \hat{\mathbf{s}}) + \gamma_2 \hat{\mathbf{s}} \times (\hat{\mathbf{r}} \times \hat{\mathbf{s}})].
    \label{eq:f_Yd}
\end{equation}
In Eq.~\eqref{eq:f_Yd}, $\hat{\mathbf{r}} = \mathbf{r}/r$ is the heliocentric unit position vector, and $\hat{\mathbf{s}}$ is the unit vector of the asteroid spin axis. In addition,
\begin{equation}
    \kappa = \frac{4 \alpha}{9} \frac{SF}{m c},
\end{equation}
where $S = \pi R^2$ is the cross section of the asteroid, $R$ is the radius, $F$ is the solar radiation flux at a heliocentric distance $r$, $\alpha$ is the absorption coefficient, $m$ is the asteroid mass, and $c$ is the speed of light. The coefficients $\gamma_1, \gamma_2$ are expressed as
\begin{equation}
    \begin{split}
        \gamma_1 & = -\frac{k_1(R'_{\textrm{d}}) \Theta_{\textrm{d}}}{1 + 2 k_2(R') \Theta_{\textrm{d}} + k_3(R'_{\textrm{d}}) \Theta_{\textrm{d}}^2},\\
        \gamma_2 & = -\frac{1 + k_2(R'_{\textrm{d}}) \Theta_{\textrm{d}}}{1 + 2k_2(R') \Theta_{\textrm{d}} + k_3(R'_{\textrm{d}}) \Theta_{\textrm{d}}^2}.
    \end{split}
    \label{eq:gamma1gamma2}
\end{equation}
In Eq.~\eqref{eq:gamma1gamma2}, $R' = R/l_{\textrm{d}}$ is the scaled radius, where $l_{\textrm{d}} = \sqrt{K/(\rho C \omega)}$ is the penetration depth of the diurnal thermal wave, obtained from the thermal conductivity $K$, the heat capacity $C$, the density $\rho$, and the rotation frequency $\omega$. The functions $k_1, k_2$, and $k_3$ are positive analytic functions of the scaled radius \citep{vokrouhlicky_1998, vokrouhlicky_1999}. In addition, $\Theta_{\textrm{d}} = \sqrt{\rho K C \omega}/(\varepsilon \sigma T_\star^3)$, where $\sigma$ is the Stefan-Boltzmann constant, and $T_\star$ is the subsolar temperature, which is given by $4\varepsilon \sigma T_\star^4 = \alpha F$.

The seasonal component is expressed as
\begin{equation}
    \mathbf{f}_{\text{Y, s}} = \kappa [ \bar{\gamma}_1 (\hat{\mathbf{r}} \cdot \hat{\mathbf{s}}) + \bar{\gamma}_2 (\hat{\mathbf{n}}\times \hat{\mathbf{r}})\cdot \hat{\mathbf{s}} ]\hat{\mathbf{s}},  
\end{equation}
where $\hat{\mathbf{n}}$ is the unit vector normal to the orbital plane. The coefficients $\bar{\gamma}_1, \bar{\gamma}_2$ have the same expressions as Eq.~\eqref{eq:gamma1gamma2}, but evaluated with
$\overline{\Theta}_{\textrm{s}} = \sqrt{\rho K C n}/(\varepsilon \sigma T_\star^3)$, and with a scaled radius $R'_{\textrm{s}} = R/l_{\textrm{s}}$, where $l_{\textrm{s}} = \sqrt{K/(\rho C n)}$ is the penetration depth of the seasonal thermal wave.
The average Yarkovsky drift $\text{d}a/\text{d}t$ is finally obtained by numerically averaging the instantaneous Yarkovsky drift of Eq.~\eqref{eq:dadt_i} over an orbital period.

\subsubsection{Modeling the orbital and physical parameters}
\label{sss:params}
The orbital parameters in the analytical formulas of Eq.~\eqref{eq:dadt_i} are the semi-major axis $a$ and the eccentricity $e$. The uncertainties on these parameters are usually small when the orbit is well determined, and they typically do not affect the results given by Eq.~\eqref{eq:dadt_i}. We therefore fixed them to their nominal value.

The other parameters in Eq.~\eqref{eq:dadt_i} are the diameter $D$, the density $\rho$, the obliquity $\gamma$, the rotation period $P$, the heat capacity $C$, the thermal conductivity $K$, the emissivity $\varepsilon$, and the absorption coefficient $\alpha$. 
The emissivity was fixed to $\varepsilon = 0.984$, which corresponds to the average value measured on meteorites \citep{ostrowsky-bryson_2019}. 
The heat capacity was fixed to a value of $C=800$ J kg$^{-1}$ K$^{-1}$, which is appropriate for NEAs \citep{delbo-etal_2015}. However, changing this value in the range $500-1200$ J kg$^{-1}$ K$^{-1}$, which is thought to be valid for most of materials present on asteroids, produces very small changes in the predicted semi-major axis drift \citep{fenucci-etal_2021, fenucci-etal_2023}.
Because we are interested in estimating the maximum semi-major axis drift that can be reached by a NEA, we assumed that the obliquity $\gamma$ can be either $0^\circ$ or $180^\circ$. For the same reason, we also fixed $\alpha = 1$.

To obtain the values of the other physical parameters, we used the \texttt{SsODNet} service\footnote{\url{https://ssp.imcce.fr/webservices/ssodnet/}} \citep{berthier-2022} through the Python \texttt{rocks} API\footnote{\url{https://github.com/maxmahlke/rocks}}. The diameter $D$, the rotation period $P$, and the taxonomic class were retrieved through \texttt{ssoCard}, which provides a best estimate by averaging all the available measurements of a given parameter. 
When the values of $D$ or $P$ were provided, we assumed that they are Gaussian distributed with a mean value equal to the nominal value and a standard deviation equal to the error.
On the other hand, the taxonomic complex gives only indirect information about the density. Table~\ref{tab:taxonomy_density} reports all the taxonomic complexes found for NEAs in the \texttt{SsODNet} database. The grouping of the different taxonomic types \citep{tholen_1984, bus-binzel_2002, deme-etal_2009} and complexes can be found in \citet{berthier-2022}. The density was assumed to be distributed according to a log-normal distribution, with values of $\mu$ and $\sigma$ provided by Table~\ref{tab:taxonomy_density}. These values were obtained by a statistics on the density values extracted from the \texttt{SsODNet} database (see Appendix~\ref{app:density}).

\begin{figure}[!ht]
    \centering
    \includegraphics[width=0.48\textwidth]{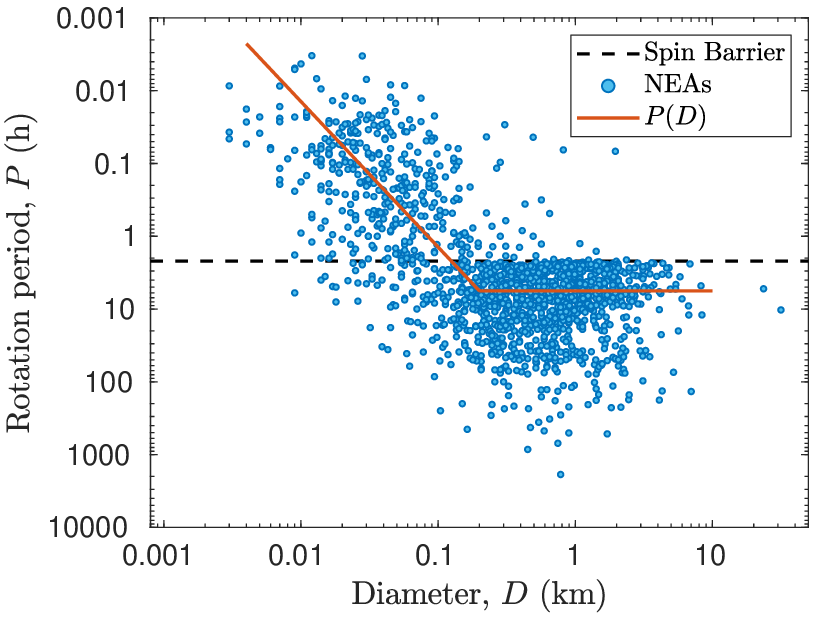}
    \caption{Distribution of the rotation period measurements of NEAs extracted from LCDB, together with the rotational disruption limit of 2.2 h and the dependence of $P(D)$ on the diameter.}
    \label{fig:P_trend}
\end{figure}

Whenever the diameter $D$, the taxonomic complex, or the rotation period $P$ were lacking in the \texttt{SsODNet} database, we modeled them according to the properties and available models of the whole NEA population.  
The distribution of the measured rotation periods shows a dependency on the
diameter \citep{pravec-harris_2000}, where objects rotating in less than about
2.2 h are found only at sizes with diameters smaller than roughly 150$-$200 m.
Here we assumed that the rotation period varied with the diameter as
\begin{equation}
    P(D) = 
    \begin{cases}
    5.6 \text{ h}                                   & \text{if } D \geq 200 \text{ m}, \\
    \displaystyle 5.6 \text{ h} \times \bigg(\frac{D}{0.2 \text{ km}}\bigg)^2 & \text{if } D < 200 \text{ m}.
    \end{cases}
    \label{eq:P_trend}
\end{equation}
Figure~\ref{fig:P_trend} shows the trend of the rotation period of Eq.~\eqref{eq:P_trend} we adopted, superimposed on the distribution of rotation period measurements of NEAs extracted from the Asteroid Lightcurve Database (LCDB) \citep{warner-etal_2009}. Above the threshold of 200 meters of diameter, we used a constant rotation period of 5.6 h, corresponding to the median value of the measured rotation periods of asteroids larger than the threshold. At sizes smaller than 200 meters, asteroids may rotate very fast without being disrupted, and therefore, we assumed a trend that scales as $D^{-2}$, which is a better fit to the data than the $D^{-1}$ trend that was previously suggested by \citet{farinella-etal_1998}. The biases in the rotation period measurements for asteroids with diameters smaller than 200 m are unclear, and it is difficult to extrapolate a general trend for the whole population. However, we found this solution to be the best that can be done based upon the currently available observations. 

For the diameter $D$ and the density $\rho$, we used the model by \citet{fenucci-etal_2021, fenucci-etal_2023}, which combines the orbital distribution model for NEAs by \citet{granvik-etal_2018} and the albedo distribution model for NEAs by \citet{morbidelli-etal_2020}. The model first produces a distribution of the visual albedo $p_V$ that is later converted into a joined distribution $(D, \rho)$ of the diameter and density. For this purpose, we also assumed that the absolute magnitude $H$ is Gaussian distributed, with a standard deviation given by the root mean square (RMS) error obtained by converting the visual magnitude measurements into absolute magnitude. 
When $D$ is given in the \texttt{SsODNet} database, but not the taxonomic complex (or when the taxonomic complex is given in the \texttt{SsODNet} database, but not the diameter $D$), we used the marginal distribution of $\rho$ (the marginal distribution of $D$) obtained by the same model.

The thermal conductivity $K$ is related to the thermal inertia $\Gamma$ by the relation
\begin{equation}
    \Gamma = \sqrt{\rho K C}.
    \label{eq:TI}
\end{equation}
We assumed $\Gamma$ to be Gaussian distributed with a mean value of 250 J m$^{-2}$ K$^{-1}$ s$^{-1/2}$ \citep{delbo-etal_2007} and a standard deviation of 100 J m$^{-2}$ K$^{-1}$ s$^{-1/2}$, and the value of $K$ was then computed by inverting Eq.~\eqref{eq:TI}.

Table~\ref{tab:phys_summary} summarizes the physical parameters for the model of the Yarkovsky effect in Eq.~\eqref{eq:dadt_i} and the corresponding value or model we used to compute the predicted drift of the semi-major axis.

\begin{table}[!ht]
    \centering
    \caption{Physical parameters for the Yarkovsky effect model and the corresponding value or model. }
    \begin{tabular}{cc}
    \hline
    \hline
    Parameter        & Value or model \\
    \hline
       $D$           & From \texttt{SsODNet} or from [1] model \\
       $\rho$        & From \texttt{SsODNet} taxonomy or from [1] model \\
       $P$           & From \texttt{SsODNet} or Eq.~\eqref{eq:P_trend} \\
       $\Gamma$      & $250 \pm 100$ J m$^{-2}$ K$^{-1}$ s$^{-1/2}$\\
       $C$           & $800$ J kg$^{-1}$ K$^{-1}$\\
       $\gamma$      & 0$^\circ$ or 180$^\circ$  \\
       $\alpha$      & 1 \\
       $\varepsilon$ & 0.984 \\
    \hline
    \end{tabular}
    \tablefoot{Reference [1] above refers to \citet{fenucci-etal_2021}. }
    \label{tab:phys_summary}
\end{table}

\subsection{Treatment of poor astrometry}
\label{ss:isol_track}
The fact that a high signal-to-noise ratio (S/N) detection of $A_2$ is found from orbit determination does not imply that it is due 
to a true dynamical effect.
In fact, previous works \citep{farnocchia-etal_2013,delvigna-etal_2018} found that old
and poorly measured astrometry or isolated tracklets that significantly
increase the observational arc \citep[see also][]{hu-etal_2023} may lead to
erroneous detections of the Yarkovsky effect.   
To treat old astrometry, we adopted data weights at 10 arcsec for measurements taken before 1890 and at 5
arcsec for those taken before 1950, so that very old observations were less
significant in the orbit determination process.

To ensure that a detection with an S/N higher than 3 is reliable and does not depend on short isolated tracklets, we adopted the following strategy.
We defined an isolated tracklet as a set of observations such that 1) they were all taken from the same observatory, 2) the length of the arc was shorter than 15 days, and 3) they were separated by at least five years from the previous and the following observation sets.
When an isolated tracklet was found, we removed it from the dataset of observations, and we repeated the process until a dataset without isolated tracklets was obtained. 
Then, we performed the 7-D OD with the dataset that did not contain isolated tracklets. When the S/N in $A_2$ was lower than 3, the detection strongly depends on the old isolated observations and may therefore not be reliable or possibly spurious. As a conservative choice, the detection was therefore discarded. 
On the contrary, if a signal with a large S/N in $A_2$ is found also with the smaller set of observations and with a possibly shorter observational arc, the signal is most likely true and due to dynamical effects rather than astrometric issues. In this case the detection is considered valid.

Note that detections are more sensitive to isolated tracklets appearing at the beginning of the observational arc, as they extend the arc length, however they could be also affected by intra-arc isolated tracklets. In addition to this, some isolated tracklets may be known to be reliable. For example, some old observations may have been carefully remeasured from an experienced observer who examined the detection and confirmed its authenticity and the correctness of the astrometry \citep[see][for an example]{vokrouhlicky-etal_2008}, or they may be produced by modern surveys using more advanced technologies. To handle these cases, we introduce a list of exceptions to the rule described above, where the isolated tracklet is always included in the orbital fit. Regarding modern surveys, we assume that only tracklets from the Pan-STARRS survey (IAU codes F51 and F52) are always included in the orbital fit, as a conservative choice.

\subsection{Automated Yarkovsky effect detection}
\label{ss:auto_yarko}
To select candidates for Yarkovsky effect detection, we first discard all the NEAs with uncertainty in semi-major axis $\sigma_a$ larger than $10^{-5}$ au, since their orbit is not known with a sufficient accuracy to attempt for the detection. Here $\sigma_a$ refers to the uncertainty of the orbit obtained from 6-dimensional orbit determination (6-D OD), estimated at the weighted mean observational epoch.
For the asteroids passing the first criterion, we compute the distribution of the predicted maximum Yarkovsky drift\footnote{Here we refer to the maximum because the obliquity is fixed at extreme values of $0^\circ$ or $180^\circ$, that maximize the diurnal Yarkovsky effect.} $(\text{d}a/\text{d}t)_{\max}$ that each of them can be subjected to. This is done with a Monte Carlo method by computing the expected Yarkovsky drift with the model of Sec.~\ref{sss:say}, using 500~000 combinations of input parameters chosen from the parameter distributions described in Sec.~\ref{sss:params}. Then, we determine the 95-th percentile of the distribution of $|(\text{d}a/\text{d}t)_{\max}|$, i.e. the value $Y_M$ such that 
\begin{equation}
    P(|(\text{d}a/\text{d}t)_{\max}| < Y_M) = 0.95,
\end{equation}
where $P(\, \cdot \,)$ denotes the probability. Values of semi-major axis drifts larger than $Y_M$ in absolute values are attained only in very unlikely combinations of the input parameters.
By denoting with $\Delta T$ the total length of the observational arc of the asteroid, if the condition
\begin{equation}
\sigma_a < \Delta T \times Y_M
\label{eq:unc_cond}
\end{equation}
is fulfilled, the asteroid is considered a good candidate for Yarkovsky effect detection, hence the value of $A_2$ is fitted through a 7-dimensional orbit determination (7-D OD). If the signal-to-noise ratio $\text{S}/\text{N} = |A_2/\sigma_{A_2}|$ is smaller than 3, then the detection is considered not statistically significant, and therefore discarded.

To understand whether a detection is spurious or if the value of $A_2$ obtained from orbit determination could be actually caused by the Yarkovsky effect, we determine whether $A_2$ is statistically compatible with the physical properties of the given object or not. To this purpose, we convert $A_2$ and its uncertainty $\sigma_{A_2}$ to semi-major axis drift values $(\text{d}a/\text{d}t)_{\text{fit}}$ and $\sigma_{(\text{d}a/\text{d}t)_{\text{fit}}}$ respectively, by using Eq.~\eqref{eq:A22dadt}. 
Then, we check if the value $(\text{d}a/\text{d}t)_{\text{fit}}$ obtained from orbit determination fulfills the condition
\begin{equation}
    |(\text{d}a/\text{d}t)_{\text{fit}}| - k \sigma_{(\text{d}a/\text{d}t)_{\text{fit}}} < Y_M,    
    \label{eq:Yarko_acceptance}
\end{equation}
where $k = 1.645$ gives the 5-th percentile of the measured Yarkovsky drift. If inequality of Eq.~\eqref{eq:Yarko_acceptance} is fulfilled, it means that semi-major axis drift values down to $|(\text{d}a/\text{d}t)_{\text{fit}}| - k \sigma_{(\text{d}a/\text{d}t)_{\text{fit}}}$ are compatible with the predictions obtained with the Yarkovsky model. On the other hand, if inequality of Eq.~\eqref{eq:Yarko_acceptance} is not valid, the value detected from astrometry is not compatible with the Yarkovsky physical model, the detection is considered spurious and therefore discarded. Note that the criterion of Eq.~\eqref{eq:Yarko_acceptance} is adopted because the distribution of the expected maximum Yarkovsky drift is not Gaussian in general. 
Finally, for those detections fulfilling Eq.~\eqref{eq:Yarko_acceptance} we check the dependence on isolated tracklets by using the method described in Sec.~\ref{ss:isol_track}, and keep only those detections passing the test.

\section{Results and discussion}
\label{s:res}

\subsection{Accepted detections}
\begin{figure}[!ht]
    \centering
    \includegraphics[width=0.48\textwidth]{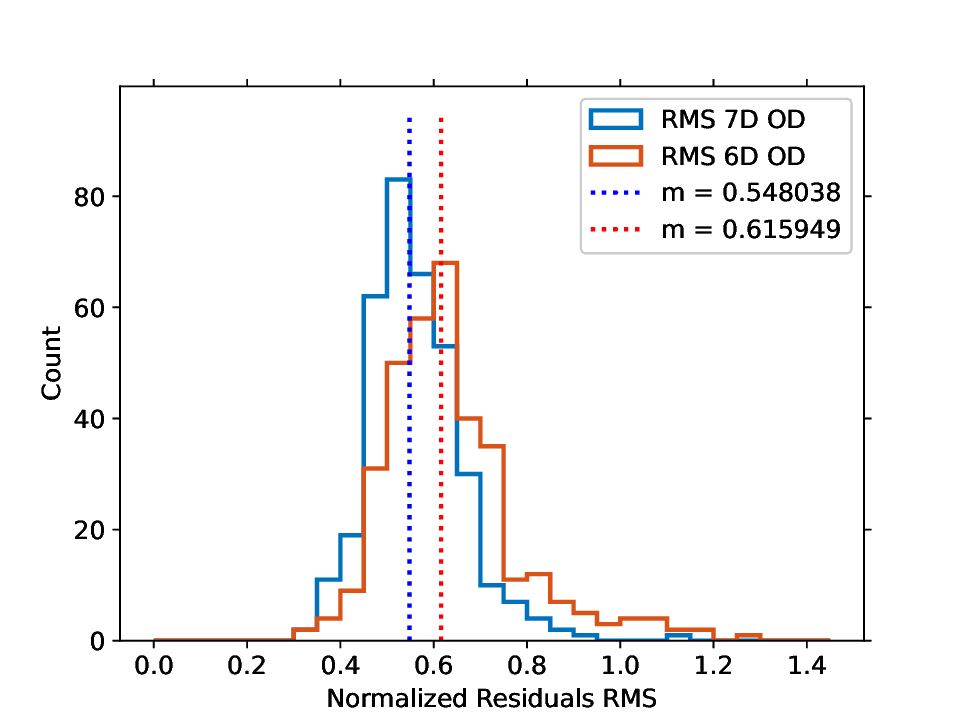}
    \caption{Distribution of the RMS of the normalized residuals for the NEAs with a positive Yarkovsy effect detection. The blue (red) histogram represents the RMS of the normalized residuals obtained with (without) the Yarkovsky effect included in the dynamical model. The dashed vertical lines correspond to the median of the two distributions, where the numerical value is reported in the legend. }
    \label{fig:RMS_hist}
\end{figure}

We ran the above procedure on the NEOCC list of known NEAs on 7 August 2023, soon after the last monthly orbital update issued by the Minor Planet Center\footnote{\url{https://www.minorplanetcenter.net/}} (MPC). Optical observations are downloaded from the MPC, while radar observations are taken from the JPL Solar System Dynamics website\footnote{\url{https://ssd.jpl.nasa.gov/sb/radar.html}}. Out of 32~396 NEAs, 5~308 were selected as candidates for a possible Yarkovsky effect detection. After the orbit determination process, we found 461 NEAs with $\text{S}/\text{N} \geq 3$, and 373 of them fulfilled also the condition of Eq.~\eqref{eq:Yarko_acceptance}. Among them, 69 objects presented isolated tracklets, and 25 were found to be spurious detections. In conclusion, we found 348 positive Yarkovsky effect detections. A short list of detections with high S/N is shown in Table~\ref{tab:det_short}, while the complete list of accepted detections in extended precision can be found at the CDS. 
Figure~\ref{fig:RMS_hist} shows the distribution of the RMS error of the normalized observational residuals of the accepted detections, obtained with 7-D OD (blue histogram). As a comparison, the red histogram shows the distribution of the RMS of normalized residuals of the same NEAs obtained with the 6-D OD. The median value decreases from 0.616 for the 6-D OD, to 0.548 for the 7-D OD. The distribution obtained with 7-D OD is skewed towards smaller values if compared to that obtained with 6-D OD, meaning that including the Yarkovsky effect in the orbital fit of these NEAs generally improves the quality of the fit of the accepted detections. 

\begin{figure}[!ht]
    \centering
    \includegraphics[width=0.48\textwidth]{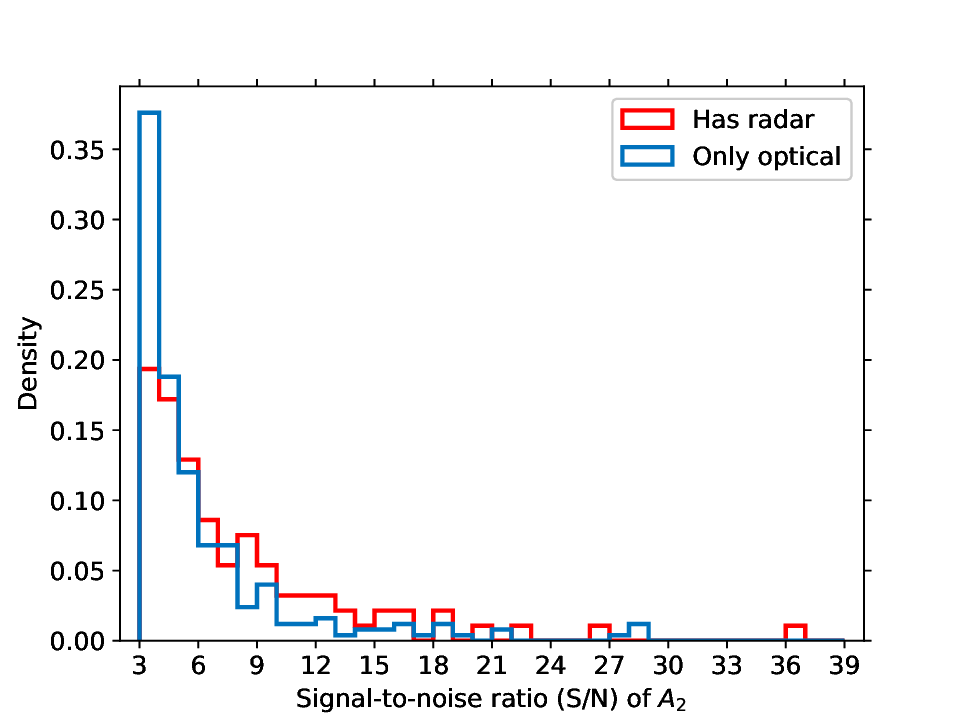}
    \caption{Distribution of the S/N of the accepted Yarkovsky effect detections, obtained with equi-spaced bins of width equal to 1. NEAs with only optical observations are represented in the blue histogram, while the red histogram refers to NEAs observed also with radar.}
    \label{fig:SNR_hist}
\end{figure}

Figure~\ref{fig:SNR_hist} shows the distribution of the S/N of the accepted Yarkovsky effect detections. About 110 objects have S/N between 3 and 4, and the presence of radar observations generally improves the S/N, even though they are not necessarily needed to obtain good detections. The asteroids with $\text{S}/\text{N} > 40$ are (101955) Bennu, 1998~SD$_9$, (99942) Apophis, (480883) 2001~YE$_4$, (524522) 2002~VE$_{68}$, (2340) Hathor, and 2002~BF$_{25}$, which are not shown in Fig.~\ref{fig:SNR_hist}. 

The Yarkovsky effect determinations can be used to extrapolate the ratio of retrograde-to-prograde rotators (R/P), because negative (positive) semi-major axis drift values are associated with retrograde (prograde) rotators. To this purpose, we randomly generate 10~000 values of $A_2$, for each positive detections. In this process, we assumed $A_2$ to be a Gaussian random variable, with mean value equal to the nominal value and standard deviation equal to the 1-$\sigma$ uncertainty of the detection. With this method, we obtained $\text{R}/\text{P} = 2.976$. Spin pole determinations of NEAs show that $\text{R}/\text{P} = 2^{+1}_{-0.7}$ \citep{laspina-etal_2004}, in good agreement from what NEA distribution models predict. Although the ratio R/P we found is still consistent with spin pole determinations, the discrepancy with the distribution models can possibly be explained by the biases in the orbital distribution of NEAs with a measured Yarkovsky effect \citep{farnocchia-etal_2013}. 
Another possible cause contributing to this discrepancy might be that the Yarkovsky detections are driven by cases with extreme obliquity, which may have a larger R/P ratio than the general NEA population.
However, it is worth noting that the percentage of retrograde rotators found here is about 75\%, which is lower than the 81\% found in \citep{farnocchia-etal_2013}, and closer to the 69\% value expected in the NEA population. This is an indication that the observation capabilities improved in the last 10 years, and the new generation surveys may help in making the sample of NEAs with measured Yarkovsky effect representative of the whole debiased NEA population. Note that a value of $\text{R}/\text{P} = 2^{+0.8}_{-0.7}$ consistent with observations was previously found in \citet{tardioli-etal_2017}, by fitting the NEA obliquity distribution to the Yarkovsky effect determinations.   

\begin{figure}[!ht]
    \centering
    \includegraphics[width=0.5\textwidth]{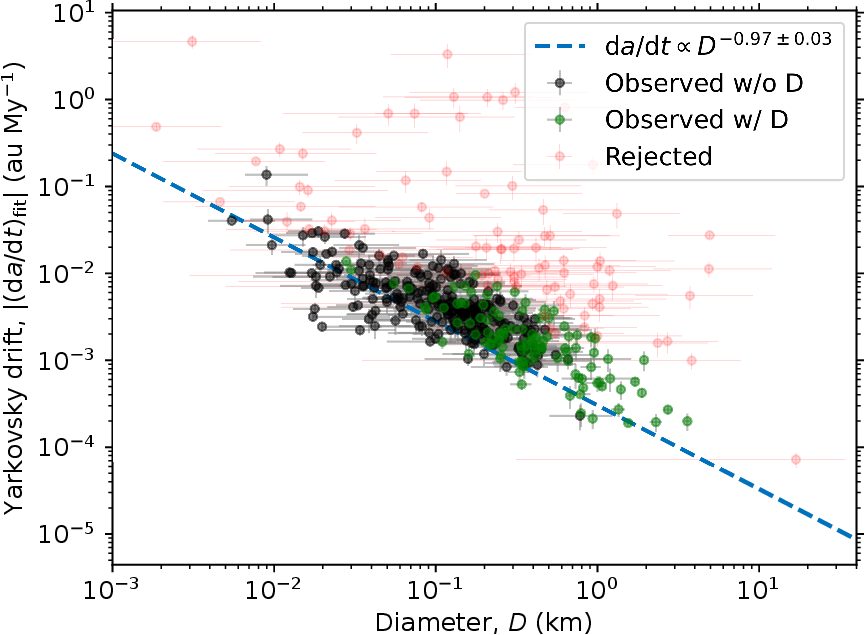}
    \caption{Distribution of the accepted Yarkovsky effect detections in the plane $(D, |(\text{d}a/\text{d}t)_{\text{fit}}|)$. Green points are NEAs for which an estimate of the diameter is available, while black points are those for which the physical model is used. The error bars in the diameter are represented by using the 15-th and 85-th percentile of the diameter distribution (see Sec.~\ref{sss:params}), which roughly correspond to the 1-$\sigma$ uncertainty in the case the diameter is estimated. The error bars in the semi-major axis drift represent the 1-$\sigma$ uncertainty obtained by the conversion of $A_2$, that has been estimated with orbit determination. The dashed blue line represents the fit with a function of the form $\alpha \times D^\beta$. Red points are detections with $\text{S}/\text{N} \geq 3$ that are considered spurious, hence rejected by our procedure.}
    \label{fig:D_vs_dadt}
\end{figure}

\begin{figure}
    \centering
    \includegraphics[width=0.5\textwidth]{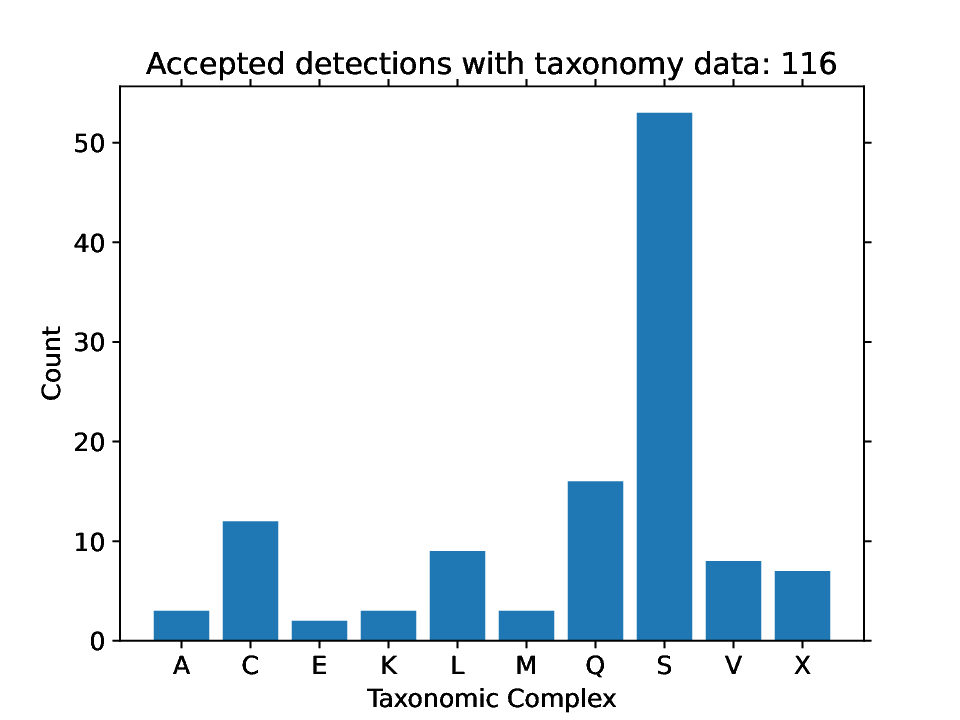}
    \caption{Taxonomy distribution of the accepted Yarkovsky effect detections. Data about the taxonomic complex were retrieved from the \texttt{SsODNet} database \citet{berthier-2022}.}
    \label{fig:count_taxonomy}
\end{figure}

Figure~\ref{fig:D_vs_dadt} shows the measured semi-major axis drifts $|(\text{d}a/\text{d}t)_{\text{fit}}|$ as a function of the diameter $D$. Error bars in the diameter were obtained by computing the 15-th and the 85-th percentile of the diameter distribution, using the model described in Sec.~\ref{sss:params}. Analytical models of the Yarkovsky effect predict that $\text{d}a/\text{d}t$ scales with the diameter as $D^{-1}$. To verify that spurious detections are correctly discarded by our procedure, we fit the data about accepted detections with a function of the form $\alpha \times D^\beta$ using the Orthogonal Distance Regression \citep[ODR;][]{boggs-etal_1987} method. The best fit values computed by the ODR algorithm are $\alpha = (31.65 \pm 2.42) \times 10^{-5}$ and $\beta = -0.97 \pm 0.03$. Note that the exponent $\beta$ is statistically compatible with the value of $-1$ predicted by Yarkovsky analytical models within 1-$\sigma$. The power law function computed with the nominal parameters is also shown in Fig.~\ref{fig:D_vs_dadt}. Detections with $\text{S}/\text{N} \geq 3$ that are considered spurious are also shown in Fig.~\ref{fig:D_vs_dadt} with red points, and most of them are far off the best-fit power law computed from the accepted detections. This already shows that our method is able to appropriately find spurious detections. 
The dispersion from the best fit power law of Fig.~\ref{fig:D_vs_dadt} is explained by the differences in the physical properties of asteroids, that are able to slightly modify the magnitude of the semi-major axis drift. To verify this we retrieved taxonomic data from the \texttt{SsODNet} database about the accepted Yarkovsky effect detections, finding 116 NEAs with such information. A histogram showing the number of NEAs with measured Yarkovsky effect for each taxonomic complex is shown in Fig.~\ref{fig:count_taxonomy}. As expected, the vast majority of the detections belong to the S-complex, however almost all the complexes (see Table~\ref{tab:taxonomy_density}) are represented in the Yarkovsky measurement sample. These facts indicate that our procedure is able to discriminate between positive and negative Yarkovsky effect detections from a statistical point of view. 

\subsection{Notable rejections}
To further show the effectiveness of the algorithm introduced here, we highlight some notable cases among the identified spurious detections. In addition to the Yarkovsky effect, other nongravitational forces such as SRP can influence the dynamics of very small bodies. So far, SRP has been detected through orbit determination only on a short list of NEAs: 2009~BD \citep{micheli-etal_2012}, 2011~MD \citep{micheli-etal_2014}, 2012~LA \citep{micheli-etal_2013}, 2012~TC$_4$, and 2015~TC$_{25}$ \citep{delvigna-etal_2018}. All these objects do not appear in our list of accepted Yarkovsky effect detections, because the semi-major axis drift obtained by the 7-D OD is not compatible with what expected from the physical model. This issue is actually fixed when SRP is added to the dynamical model, and the area-to-mass ratio $A/m$ of the object is added to the list of parameters to be estimated in the orbital fit. 

A similar result has been found for 2021~GM$_1$, an NEA that is currently in the NEOCC Risk List\footnote{\url{https://neo.ssa.esa.int/risk-list}}. The Yarkovsky effect detection of this object is rejected by our procedure, but a nongravitational effect is needed to better fit the observations. Adding SRP to the dynamical model, we fitted an area-to-mass ratio of $0.2564 \pm 0.00257$ kg$^2$ t$^{-1}$ (see Table~\ref{tab:2021GM1_orbit} for the complete orbital solution), providing a new SRP detection that was not available in the literature\footnote{We note that this detection was however already present in the JPL SBDB since May 2021.}. This shows that the procedure may also be helpful in finding asteroid candidates for detection of other nongravitational effects.
\begin{table}[!ht]
        \renewcommand{\arraystretch}{1.1}
    \centering
    \caption{Orbital elements of 2021 GM$_{1}$.}
    \tiny
    \begin{tabular}{lcl}
        \hline
        \hline
        Six-parameter solution & & Unit \\
        \hline
        Semi-major axis $(a)$             & $0.968020575 \pm 1.43\times10^{-8}$ & au \\
        Eccentricity $(e)$                & $0.041997691   \pm 1.58\times10^{-8}$ & / \\
        Inclination $(i)$                 & $1.598736270      \pm 4.09\times10^{-7}$ & deg \\
        Long. of Asc. Node $(\Omega)$     & $193.321099298    \pm 1.95\times10^{-6}$ & deg \\
        Argument of Perihelion $(\omega)$ & $207.383938594    \pm 8.52\times10^{-6}$ & deg \\
        Mean anomaly $(M)$                & $162.303893236    \pm 1.00\times10^{-5}$ & deg \\ 
        Normalized residuals RMS                     & $0.776$                                     & / \\
        \hline
        Seven-parameter solution & \\
        \hline
        Semi-major axis $(a)$             & $0.968020505 \pm 1.38\times10^{-8}$ & au \\
        Eccentricity $(e)$                & $0.041997755   \pm 1.49\times10^{-8}$ & / \\
        Inclination $(i)$                 & $1.598740133     \pm 5.18\times10^{-7}$ & deg \\
        Long. of Asc. Node $(\Omega)$     & $193.321107921   \pm 2.15\times10^{-6}$ & deg \\
        Argument of Perihelion $(\omega)$ & $207.383858669   \pm 1.14\times10^{-5}$ & deg \\
        Mean anomaly $(M)$                & $162.303967843   \pm 1.23\times10^{-5}$ & deg \\ 
        Area-to-mass $(A/m)$              & $0.2564    \pm 0.0257$ & m$^2$ t$^{-1}$\\
        Normalized residuals RMS                     & $0.622$                                     & / \\
        \hline
    \end{tabular}
    \tablefoot{Orbital elements refer to date 2021$-$04$-$14.660191 (MJD 59318.660191 TDT). Two solutions are shown: one determined with 6-D OD, and another one with 7-OD OD using SRP as additional force.}
    \label{tab:2021GM1_orbit}
\end{table}

Other interesting cases are 2006~RH$_{120}$ \citep{seligman-etal_2023} and (523599) 2003~RM \citep{farnocchia-etal_2023}. In addition to the radial component $A_1$ and the tangential component $A_2$, it is possible to also fit a nongravitational out-of-plane force component $A_3$ for these two NEAs, showing a dynamical behavior similar to that of comets. In these two cases, the value of $A_2$ obtained by a 7-D OD only is not compatible with the Yarkovsky physical model (see also Table~\ref{tab:jpl_comparison}), while it is likely due to outgassing \citep{chesley-etal_2016, taylor-etal_2023}. Still, these detections are recognized as spurious by our procedure. 

Another remarkable detection that was flagged as spurious is (433) Eros. This NEA was selected as a candidate for Yarkovsky effect detection because of its long observational arc of more than 100 years, and $A_2$ was fitted with a S/N of about 5. However, the detection was not statistically compatible with the prediction made by the Yarkovsky physical model, and therefore the detection is spurious. This detection needs to be necessarily discarded, because bad quality astrometric measurements performed before the 1950 cause an artificial nongravitational acceleration that is actually not expected. In fact, (433) Eros is known to have an obliquity of almost $90^\circ$ \citep{yeomans-etal_2000}, that minimizes the Yarvkosky effect, but the procedure is able to discard the detection even by assuming an obliquity of $0^\circ$ or $180^\circ$ (see Sec.~\ref{sss:params}).    
All these considerations provide evidences of the effectiveness of the algorithm introduced here, in both the selection of possible candidates for the Yarkovsky effect determination, and in the identification of spurious detections. 

The algorithm found 69 $A_2$ detections with S/N larger than 3 and fulfilling Eq.~\eqref{eq:Yarko_acceptance} that were affected by the presence of isolated tracklets in their dataset. Among them, 58 presented isolated tracklets at the beginning of the observational arc, and the complete list of these asteroids is reported in Table~\ref{tab:validation_final}. The table reports also the S/N of $A_2$ obtained without including the isolated tracklet in the orbital fit, following the procedure described in Sec.~\ref{ss:isol_track}, and 30 of them still present a signal in $A_2$ even with the shorter observational arc. These objects were therefore accepted as positive Yarkovsky effect detections. Regarding the list of exceptions for which we know that astrometry is reliable, up to now it includes: 2012~BB$_{124}$ and 2014~RO$_{17}$, for which accurate measurements were performed by MM; 2015~LA$_2$, 2018~KR, and 2016~UW$_{80}$, that present isolated tracklets from F51; (152563) 1992~BF for which it is known that the tracklet has been accurately remeasured \citep{delvigna-etal_2018}. While this list may increase in the future, it is not expected to grow to a point that can not be maintained manually. 
Other 11 objects presented intra-arc isolated tracklets, but the $A_2$ detection was affected only for 2005~TG$_{50}$ and 2011~XC, that were therefore considered as spurious. In addition, the S/N drops down also for 2009~OS$_5$ when observations from 2014 from Mauna Kea Observatory (IAU code 568) are excluded in the fit, however we considered them as reliable\footnote{These observations have program code equal to 2, corresponding to D. Tholen.} and we therefore accepted this detection.
Finally, among the detections that were not accepted, we note that the algorithm flagged (469219) Kamo'oalewa as spurious, in agreement with \cite{hu-etal_2023}.

\subsection{Publication on the NEOCC portal}
The availability of a large number of reliable Yarkovsky effect detections opens possible future developments in other fields of asteroid research, such as physical properties determination \citep{rozitis-green_2014, fenucci-etal_2021, fenucci-etal_2023}, long term dynamical studies \citep{fenucci-novakovic_2021}, or population studies on NEAs \citep{tardioli-etal_2017}. To this end, it is important to continuously provide up-to-date data to the scientific community. The procedure described here has been now adopted by the ESA NEOCC to periodically update the Yarkovsky effect detections on NEAs, that are automatically posted on the NEOCC web portal\footnote{\url{https://neo.ssa.esa.int/}} and available through the HTTPS APIs\footnote{\url{https://neo.ssa.esa.int/computer-access}}. The catalogues of observations and of known NEAs are continuously growing, and the algorithm is scheduled to run after every monthly orbital update issued by the MPC, so that old determinations are updated and possible new detections are added to the database. 

\subsection{Comparison with the JPL SBDB}
\begin{figure}[!ht]
    \centering
    \includegraphics[width=0.5\textwidth]{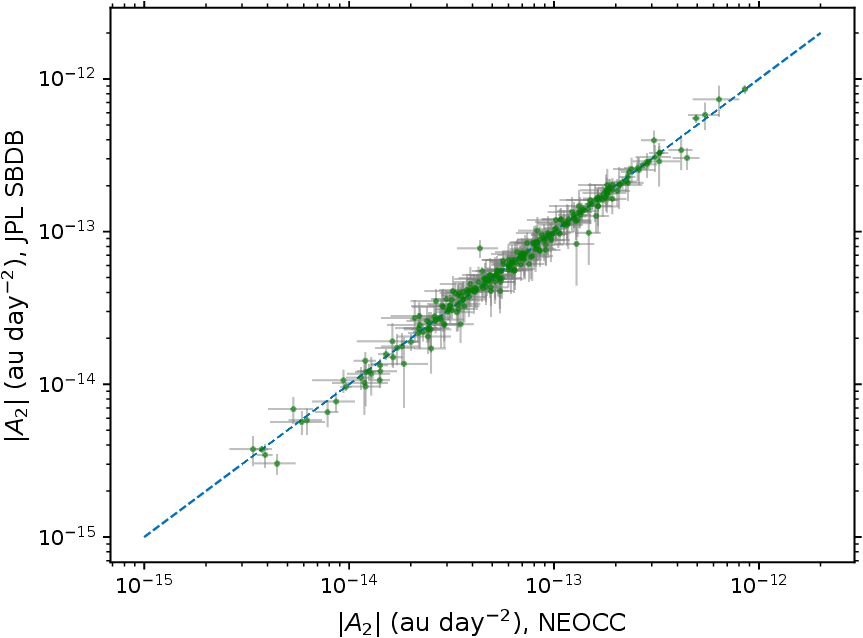}
    \caption{Distribution of the common $A_2$ measurements found in this work and at the JPL SBDB. The blue dashed line is the diagonal.}
    \label{fig:NEOCC_vs_JPL}
\end{figure}

We downloaded the list of determinations of $A_2$ from the JPL Small Body Database\footnote{\url{https://ssd.jpl.nasa.gov/}} (SBDB) maintained by the JPL SSD group on 8 August 2023. A total of 333 measurements, 308 of which with $\text{S}/\text{N} \geq 3$, were found. 
Differences in the number of detections with a good S/N are possibly due to different debiasing and weighting schemes of observations, different observational datasets and rejections, and timing errors handling. These differences become particularly significant when the S/N is close to the threshold of 3.
Among the JPL detections, 13 objects have also a determination of the component $A_1$, and 6 objects have a determination of both $A_1$ and $A_3$. We did not make the comparison for these objects, because the results are obtained with different dynamical models. 
Of the remaining objects, 270 were found in our list of accepted Yarkovsky effect measurements. Figure~\ref{fig:NEOCC_vs_JPL} shows the scatter plot of the common $A_2$ determinations, together with their 1-$\sigma$ uncertainty. We fitted the data with a linear function of the form $\alpha |A_2| + \beta$ with the ODR algorithm, that gave an estimation of the parameters of $\alpha = 1.00 \pm 0.0001$ and $\beta = (-1.90 \pm 0.97) \times 10^{-16}$, meaning that our determinations and those of the JPL SBDB are statistically compatible, and no clear outlier can be recognized. This is certainly due to the fact that the orbit determination algorithms and the datasets used by the JPL SSD group are similar to those used here for these cases.
To make a quantitative comparison of the results for each asteroid, we computed the relative errors 
\begin{equation}
    \varepsilon_1 = \frac{|A_2 - A_2^{\text{JPL}}|}{\sigma_{A_2}}, \quad \varepsilon_2 = \frac{|A_2 - A_2^{\text{JPL}}|}{\sigma_{A^{\text{JPL}}_2}},
    \label{eq:A2_relerr}
\end{equation}
where the superscript JPL denotes the value provided by the JPL SBDB. For the vast majority of the asteroids in common, the relative errors $\varepsilon_1, \varepsilon_2$ were both smaller than 3, indicating that the results provided by us and by the JPL SBDB are all statistically comparable, even case by case. The only cases in which the relative errors are larger than 3 are (65717) 1993 BX$_3$, 2003~AF$_{23}$, (1685) Toro. However, also in these cases the values of $\varepsilon_1$ and $\varepsilon_2$ are only slightly larger than 3. We also performed a test by checking that
\begin{equation}
    \chi = \frac{|A_2 - A_2^{\text{JPL}}|}{\sqrt{(\sigma_{A_2})^2 + (\sigma_{A^{\text{JPL}}_2})^2}} < 2,
\end{equation}
meaning that the hypothesis that the two distributions are not the same is rejected at 95\% confidence level. For all the common asteroids with a Yarkovsky effect measurement we found $\chi < 2$, with the only exceptions of 2003 AF$_{23}$ and (65717) 1993 BX$_3$, that have $\chi = 2.23$ and $\chi = 2.18$, respectively. However, the two measurements are still considered the same if the test is verified with a more relaxed confidence level of 90\%. In addition, we found that the vast majority of the measurements have $\chi<1$, implying an excellent agreement between our results are those reported in the JPL SBDB.
Table~\ref{tab:chivalues} reports the numerical values of the relative errors $\varepsilon_1$ and $\varepsilon_2$, and of $\chi$ for all the NEAs with $\chi >1$. The full table in extended precision, used to generate Fig.~\ref{fig:NEOCC_vs_JPL}, is available through the CDS. 

Asteroids with a determination of $A_2$ that are reported in the JPL SBDB, but are either discarded or not found by our procedure, are 54. Among them, 44 have $\textrm{S}/\textrm{N} \geq 3$ and their list is shown in Table~\ref{tab:jpl_comparison}, while the complete list in extended precision is available through the CDS.
%
Notable cases are those of (65803) Didymos, (3200) Phaethon, and (25143) Itokawa. 

For asteroid (65803) Didymos the JPL SBDB reports an orbital solution with S/N $= 16$ in the $A_2$ parameter, while the value we obtained here is only of 2.57. The reasons for this discrepancy are probably due to the availability of some occultation observations took during the September 2022 campaign \citep{dunham-etal_2023}, the availability of data from the Double Asteroid Redirection Test \citep[DART,][]{cheng-etal_2018} mission, and the inclusion of Gaia astrometry in the orbit computation. These high accuracy observations are not reported by the MPC, and therefore they were not considered in our data set. 

For (3200) Phaethon the JPL SBDB reports an orbital fit with a S/N in $A_2$ of 8.3. The Yarkovsky effect measurement for this object has been reported by \citet{hanus-etal_2018}, where the authors included also very precise Gaia astrometric observations taken between 2014 and 2016 in the dataset. After the publication of that work, occultation observations taken on 25 October 2019 permitted to further improve the orbit of (3200) Phaethon \citep{dunham-etal_2021}. However, both the Gaia astrometry and the occultations are currently not reported by the MPC, and we believe this being the reason we are not able to fit the $A_2$ parameter with such high accuracy.  

The last case is that of (25143) Itokawa, for which the JPL SBDB gives an orbital solution with S/N of 3.2 in the $A_2$ parameter. In this case, it is not easy to identify what may cause this discrepancy in the results, and the only noticeable difference that we could find is in the number of observations used in the orbital fit.
For other objects, we also obtain a S/N larger than 3 as for the JPL SBDB, but they are discarded because the drift is too large to be explained by the Yarkovsky effect, and therefore Eq.~\eqref{eq:Yarko_acceptance} does not hold.

\subsection{Limitations of the procedure}
A possible limitation of the procedure may be due to the fact that the Yarkovsky effect modeling takes into account the orbital elements of the NEA at a certain epoch to compute the expected drift $\text{d}a/\text{d}t$. However, this may have consequences if an object has a close encounter with a planet that is able to change its orbital elements. While this is certainly true, the changes induced in the 95-th percentile $Y_M$ are negligible because the distribution of the expected semi-major axis drift is mostly controlled by the uncertainties in the physical properties. To show this, we took into account the case of (99942) Apophis, for which we have an estimate of the diameter, a measurement of the rotation period, and we know that the spectral type is S. Therefore, the physical model of (99942) Apophis has certainly more constraints than the generic NEA. To show that changes in the orbital elements have a little influence on the results, we computed $Y_M$ with the orbital elements computed at the mean epoch, and with those obtained by propagating the nominal orbit to an epoch right after the close approach of 13 April 2029, that changes the semi-major axis by about 0.18 au. The value of $Y_M$ went from $22.45 \times 10^{-4}$ au My$^{-1}$ obtained before the close approach, to $22.90 \times 10^{-4}$ au My$^{-1}$ computed after the close approach. For the generic NEA we expect the change to be even smaller, because the distribution of the expected maximum semi-major axis drift is dominated by the uncertainties in the physical properties, rather than by the uncertainties in the orbital elements.

From the orbit determination point of view, an effect that is not taken into account in the model described in Sec.~\ref{ss:OD} is that of timing errors in the astrometric observations. It has been shown that the astrometric errors are affected by time calibration errors made by observers. These calibration errors can be either random or systematic, and they are especially important when an object has a fast apparent motion \citep{farnocchia-etal_2022}. In turn, this may result in an erroneous estimation of the $A_2$ parameter, thus affecting the Yarkovsky effect detection. In this respect, the Aegis software used at the NEOCC is already able to treat timing errors by correcting the weights of the observations using the correlation with the right ascension and declination, with the method introduced in \citet{farnocchia-etal_2022}. However, the information about the timing errors is communicated only in the new Astrometry Data Exchange Standard \citep[ADES,][]{chesley-etal_2017}, which is not currently used to automatically compute asteroid orbits at the NEOCC \citep{faggioli-etal_2023}. The switch of the Aegis system to automatically work with observations in ADES format is planned for the future, and the procedure described here will be able to automatically take into account also timing errors in the orbit determination.

Finally, the detection of a nongravitational force $A_2$ along the transverse direction does not necessarily mean that the Yarkovsky effect is responsible for it. In fact, other dynamical effects, such as outgassing or mass-shedding, may produce a transversal acceleration \citep{farnocchia-etal_2023}. The usage of a statistical model for the physical properties of NEAs seems to help identifying these outliers, as confirmed by the cases of (523599) 2003 RM and 2006 RH$_{120}$ that were flagged as spurious. Still, there may be cases in which the acceleration produced by such phenomena is small enough to result compatible with the acceleration produced by the Yarkovsky effect. In these cases, the detection of such acceleration is real, and it is not possible to make a clear distinction between the Yarkovsky effect and activity events, unless there are proofs of such phenomena obtained from detailed studies of the single object.  

\section{Conclusions}
\label{s:conclusions}
In this paper we introduced a procedure for the automated Yarkovsky effect detection on NEAs and for the identification of spurious detections. The procedure makes use of a statistical model to estimate the physical properties of an NEA, based on the most recent data and models about the NEA population. Physical properties are then used to estimate the expected maximum Yarkovsky semi-major axis drift. Candidates for detection are selected according to the current precision in the knowledge of the orbit, the observational arc, and the expected semi-major axis drift. Then, a 7-dimensional orbit determination of the candidates, that includes the estimation of a transversal nongravitational component, is performed. 
If the S/N of the $A_2$ detection is larger than 3, a statistical test to check the compatibility with the Yarkovsky effect physical model is carried out. Finally, the dependence of the detection on isolated tracklets is analyzed.

Among the known NEAs, we identified 348 positive Yarkovsky effect detections. This procedure is now adopted by the ESA NEO Coordination Centre, and data are automatically uploaded on the web portal and updated at every monthly orbital update issued by the MPC. A comparison with the results obtained by the JPL SSD is also shown, and the measurements are generally in a very good agreement, being based on similar orbit determination methods and datasets.

\begin{acknowledgements}
We thank Davide Farnocchia for the constructive comments on an early version of the manuscript, that permitted us to improve the quality of the work.
\end{acknowledgements}

\bibliographystyle{aa}
\bibliography{holyBib} 

\begin{appendix}

\section{Densities of asteroid complexes}
\label{app:density}
To determine values of the densities of asteroid complexes, we extracted data from the \texttt{SsODNet} database using the best value provided by the \texttt{ssoCard}. We selected only asteroids for which the taxonomic complex and the density were both determined, and we found a total of 386 objects. To determine the density distribution of a complex, we used the following procedure: 1) the density distribution of each asteroid is assumed to be log-normal; 2) a density sample for all the asteroids belonging to a given complex was randomly generated, obtaining a sample for the whole complex; 3) the sample for the whole complex was fitted with a log-normal distribution. 
\begin{table}[ht]
    \setlength{\tabcolsep}{2pt} 
    \centering
    \caption{Taxonomic complexes included in the \texttt{SsODNet} database.}
    \begin{tabular}{cccccccc}
        \hline
        \hline
        Tax. & Tot. & \# S/N$>$3 &\#S/N$>$2 &\#S/N$>$1.5 & $\mu$ (kg m$^{-3}$) & $\sigma$ (kg m$^{-3}$)  \\
         \hline
         A   &  2   &  2   &   2    &  2    &  N/A      &  N/A    \\
         B   &  5   &  3   &   4    &  4    &  1290     &  181  \\
         C   &  78  &  30  &   44   &  72   &  1757     &  1088 \\ 
         Ch  &  106 &  33  &   53   &  88   &  1739     &  1054 \\
         D   &  4   &  1   &   1    &  1    &  N/A      &  N/A    \\
         E   &  5   &  3   &   3    &  3    &  2560     &  270  \\ 
         K   &  3   &  1   &   2    &  2    &  N/A      &  N/A    \\
         L   &  3   &  0   &   0    &  2    &  2674     &  1551 \\
         M   &  27  &  10  &   12   &  19   &  3614     &  1113 \\
         P   &  54  &  14  &   20   &  40   &  N/A      &  N/A    \\
         Q   &  3   &  3   &   3    &  3    &  1915     &  932  \\ 
         R   &  1   &  0   &   0    &  1    &  N/A      &  N/A    \\
         S   &  80  &  37  &   44   &  68   &  2721     &  1273 \\
         U   &  1   &  1   &   1    &  1    &  N/A      &  N/A    \\
         V   &  5   &  4   &   4    &  5    &  1930     &  1000 \\ 
         X   &  5   &  1   &   1    &  4    &  1427     &  1494 \\
         Z   &  4   &  0   &   1    &  3    &  2225     &  1804 \\
        \hline
    \end{tabular}
    \tablefoot{Columns report the name of the complex, the number of asteroids in \texttt{SsODNet} for which the density was determined, the number of determinations with S/N larger than 3, 2, and 1.5, and the values of the density assumed in this work. The values of $\mu$ and $\sigma$ for the log-normal distribution were obtained by fitting the available data. The flag N/A indicates the complexes for which the data were not enough to obtain a reasonable fit.}
    \label{tab:taxonomy_density}
\end{table}
\begin{figure*}[!ht]
    \centering
    \includegraphics[width=0.48\textwidth]{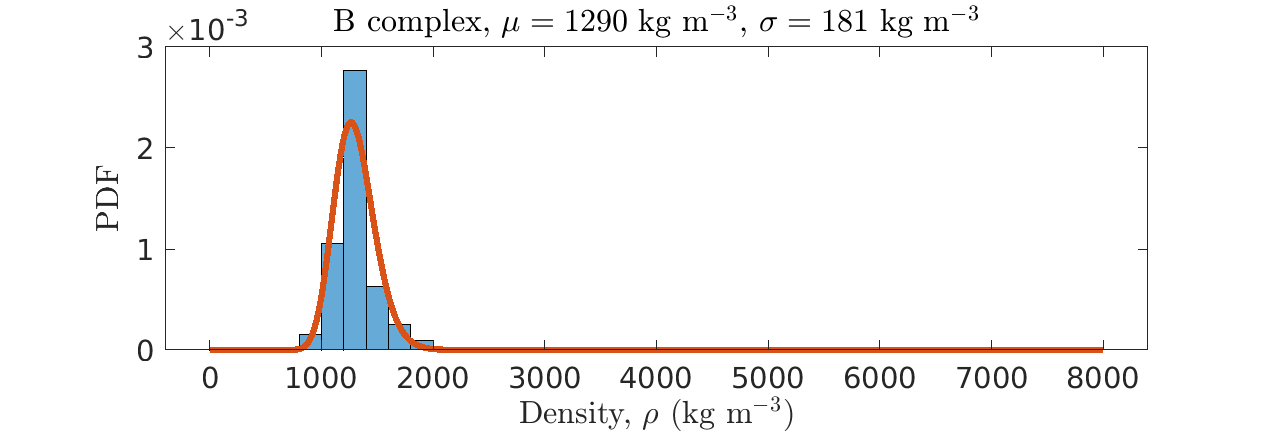}
    \includegraphics[width=0.48\textwidth]{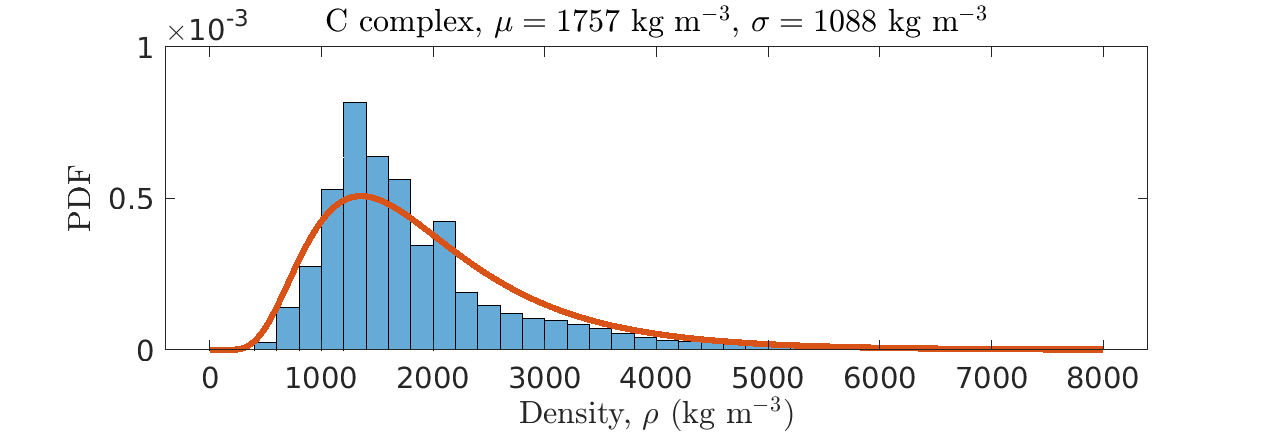}
    \includegraphics[width=0.48\textwidth]{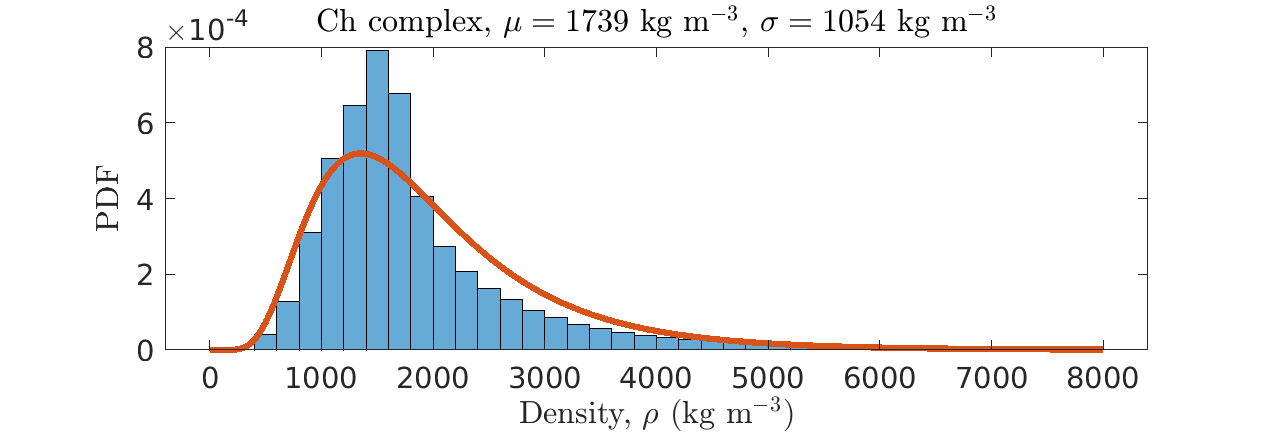}
    \includegraphics[width=0.48\textwidth]{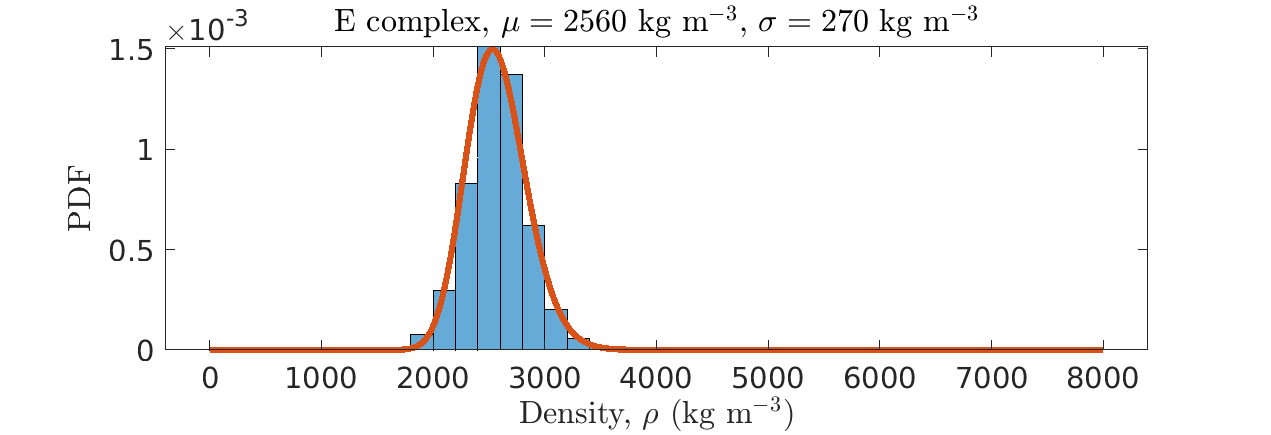}
    \includegraphics[width=0.48\textwidth]{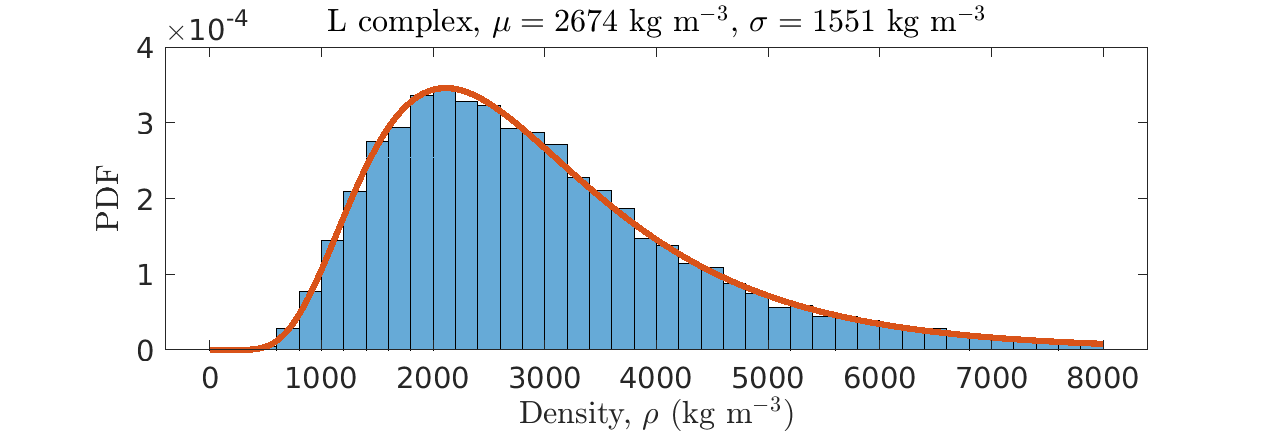}
    \includegraphics[width=0.48\textwidth]{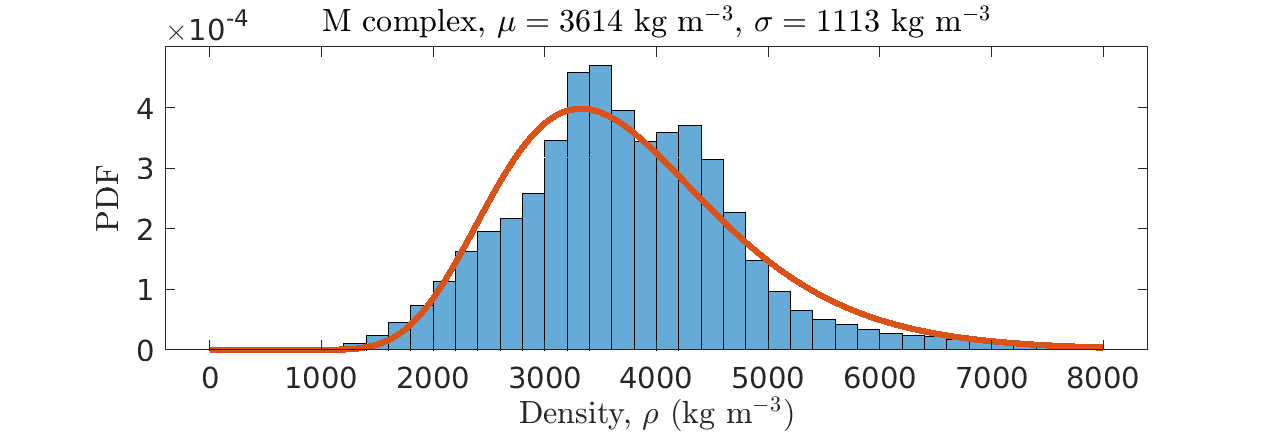}
    \includegraphics[width=0.48\textwidth]{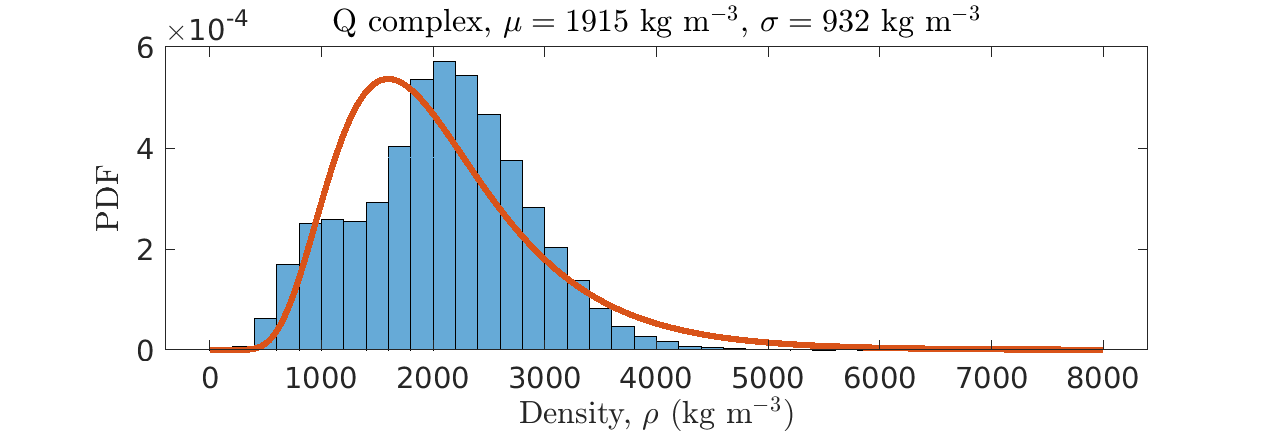}
    \includegraphics[width=0.48\textwidth]{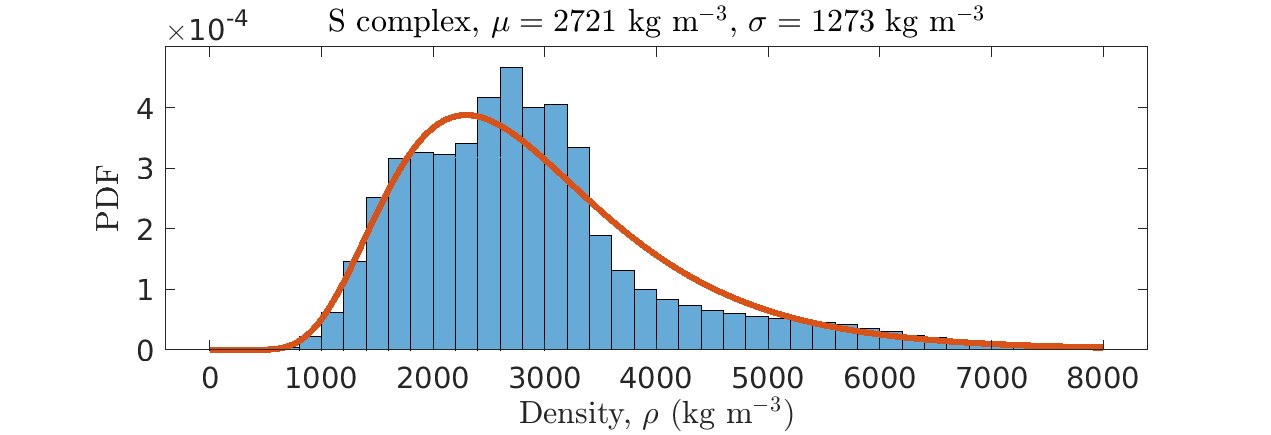}
    \includegraphics[width=0.48\textwidth]{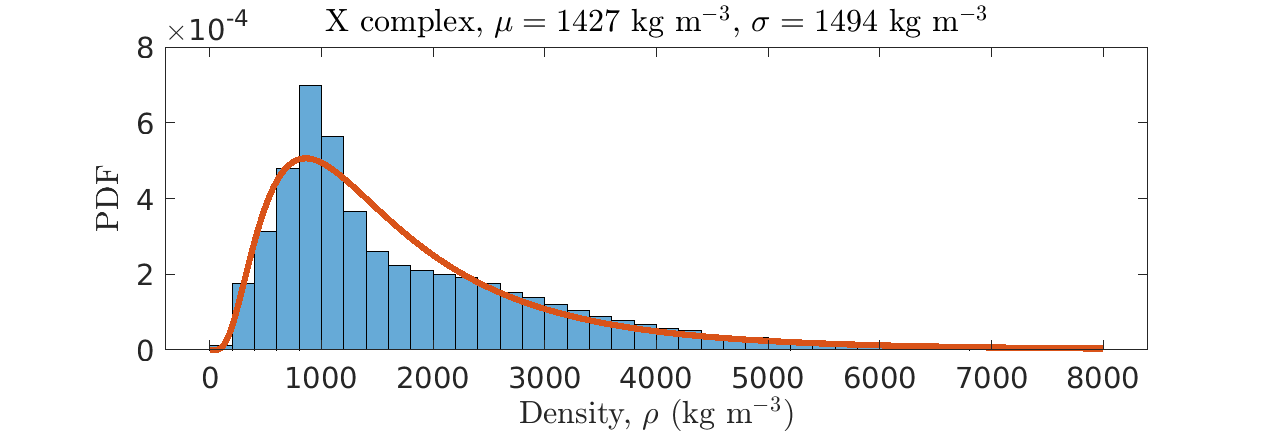}
    \includegraphics[width=0.48\textwidth]{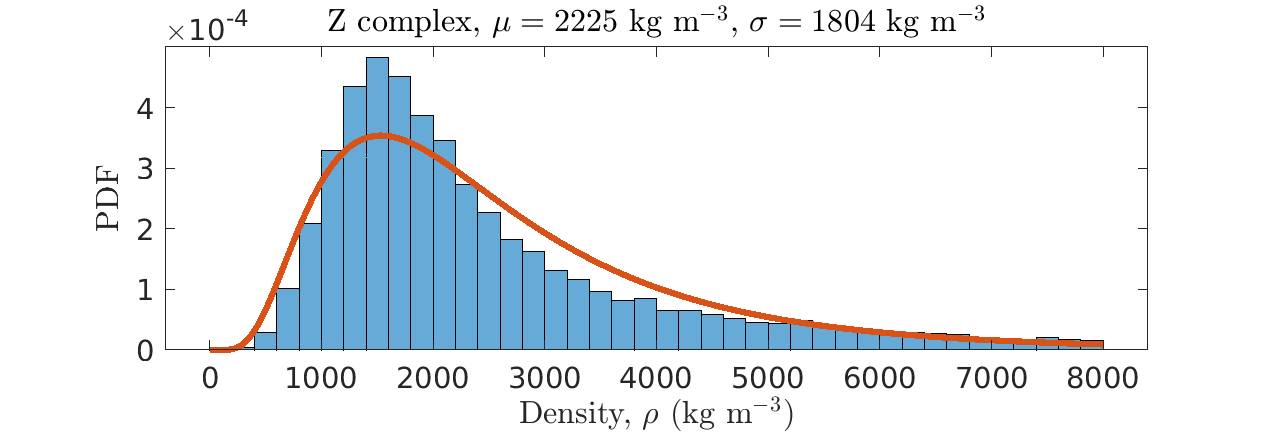}
    \caption{Fit of the density distribution for the B$-$, C$-$, Ch$-$, E$-$, L$-$, M$-$, Q$-$, S$-$, X$-$, and Z$-$complexes. The histogram is obtained by generating log-normal distributions of asteroid density determinations of the corresponding complex, while the red curve is the best-fit log-normal distribution. Numerical values of the parameters $\mu$ and $\sigma$ of the best-fit distribution are shown in the title of each panel.}
    \label{fig:density_fits}
\end{figure*}

Table~\ref{tab:taxonomy_density} shows the list of asteroid complexes found in the \texttt{SsODNet}, together with the total number of density determinations, and the number of determinations with signal-to-noise ratio (S/N) larger than 3, 2, and 1.5. The sixth and seventh columns contain the value adopted in this work for each complex. The values for the C$-$, Ch$-$, M$-$, Q$-$, and S$-$complexes were fitted by using only the data with $\text{S}/\text{N} > 3$. The values for the L$-$, X$-$, and Z$-$complexes were obtained from the measurements with $\text{S}/\text{N} > 1.5$. 
The B$-$complex showed two outliers with relatively high density (i.e. (2) Pallas and (426) Hippo), that we removed from the sample used for the density fit.
The E$-$complex also has a similar issue, because asteroids (44) Nysa and (3671) Dionysus show a density that is considered too small for these asteroid types, and therefore we removed them from the final fit. 
For the A$-$, P$-$, R$-$, K$-$, and U$-$complexes it was not possible to reasonably fit the sample with a log-normal distribution, therefore we used the population model mentioned in Sec.~\ref{sss:params}.
Finally, the densities determined for asteroids in the V$-$complex show a large variety of values, therefore we adopted the value reported by \citet{carry_2012}. The log-normal fits obtained for the with the method described above for the B$-$, C$-$, Ch$-$, E$-$, L$-$, M$-$, Q$-$, S$-$, X$-$, and Z$-$complexes are all shown in Fig.~\ref{fig:density_fits}

\section{Additional tables}

\begin{table*}[hbt!]
    \setlength{\tabcolsep}{1.5pt} 
        \renewcommand{\arraystretch}{1.3}
    \centering
    \caption{Results of the automated Yarkovsky effect determination procedure.}
    \tiny
    \begin{tabular}{ccccrrrrrrrrrrrrrrrrrrr}
    \hline
\hline
Ast.  &        $H$ &       RMS &     RMS$_6$ &            $A_2$ &    $\sigma(A_2)$ &          $\text{d}a/d\text{t}$ &    $\sigma(\text{d}a/d\text{t})$ &    $Y_M$ &         S/N &  Res. & \#OO &   \#ReO  &   \#ReO$_6$  &  \#RaO  &   \#ReRa  &   \#ReRa$_6$  &    $D_{15}$  &     $D_{50}$  &      $D_{85}$  &   $P$  & Tax   & $\Delta T$   \\
\hline
101955    &  20.6  & 0.40  & 0.45 &  $-$4.52  & 0.023  &  $-$18.89   &    0.009   &  23.20   &  191      & Acc.        &           580  &           8   &         8   &          29  &          0   &            3      &   469        &    488 &   507          &   4.29  & B     &     18.6   \\
1998 SD$_9$   &  23.9  & 0.78  & 6.84 & $-$28.98  & 0.156  &  $-$197.06  &    1.063   &  317.9   & 185       & Acc.        &           149  &           4   &         5   &          1   &          0   &            0      &   25         &    35  &   75           &   /     & /     &    19.9    \\
99942     &  18.9  & 0.30  & 0.33 & $-$2.88   & 0.023  &  $-$13.25   &    0.107   &  28.10   & 122       & Acc.        &          9523  &          18   &        20   &         50   &          0   &            5      &  300         &   340  &  380           &   27.38 & S     &    17.1     \\
480883    &  20.9    & 0.47  & 2.48 & $-$6.91   & 0.073  &  $-$50.44   &    0.533   &  116.11  &  94       & Acc.        &           430  &           1   &         5   &          7   &          0   &            2      &  103         &   172  &  287           &   /     & V     &    20.0    \\
2340      &  20.4  & 0.72  & 0.75 & $-$2.99   & 0.052  &  $-$17.38   &    0.304   &  80.28   &  57       & Acc.        &          1369  &          35   &        78   &          7   &          0   &            0      &  134         &   193  &  312           &    3.35 & Q     &    45.1     \\
524522    &  20.5   & 0.46  & 1.09 & $-$5.98   & 0.105  &  $-$35.91   &    0.630   &  42.57   &  56       & Acc.        &           606  &           8   &        21   &         10   &          0   &            7      &  112         &   176  &  337           &   13.43 & /     &    16.0  \\
2002 BF$_{25}$  &  22.4   & 0.45  & 0.88 &  16.67  & 0.295  &   71.90   &    0.0001 &  75.61   &  56       & Acc.        &           143  &           0   &        48   &          4   &          0   &            0      &  149         &   152  &  155           &   /     & /     &    20.0  \\
2012 UR$_{158}$ &  20.6  & 0.57  & 0.63 & $-$771.30 & 17.221 &  $-$8147.57 &    181.918 &  63.39   &  44       & Rej.        &           161  &           1   &        15   &          0   &          0   &            0      &  320         &   629  &  938           &   /     & /     &    10.0  \\
523599    &  19.7  & 0.46  & 1.12 &  212.06 & 5.103  &   829.24  &    19.957  &  21.83   &  41       & Rej.        &           329  &           4   &        51   &          0   &          0   &            0      &   66         &   200  &  602           &   /     & /     &    15.2  \\
2000 EB$_{14}$  &  23.2  & 0.58  & 1.04 & $-$27.00  & 0.685  &  $-$160.60  &    4.0749  &  214.11  &  39       & Acc.        &            83  &           2   &        27   &          0   &          0   &            0      &   38         &    55  &   97           &    1.79 & /     &    22.0  \\
483656    &  23.7  & 0.60  & 1.25 & $-$14.01  & 0.383  &  $-$80.064  &    2.1925  &  267.66  &  36       & Acc.        &           207  &           5   &        41   &          1   &          0   &            0      &   28         &    40  &   84           &    1.69 & /     &    18.0  \\
2009 BD    &  28.2  & 0.44  & 0.96 & $-$157.65 & 4.955  &  $-$667.46  &    20.979  &  285.39  &  31       & Rej.        &           197  &           0   &         1   &          0   &          0   &            0      &    2         &     4  &    8           &   /     & /     &     2.4  \\
\hline
    \end{tabular}
        \tablefoot{
    Columns are: Ast. = Designation of the asteroid;
    $H$ = Absolute magnitude;
    RMS = RMS of normalized residuals for 7-D OD;
    RMS$_6$ = RMS of normalized residuals for 6-D OD;
    $A_2$ = Transversal nongravitational component obtained with 7-D OD;
    $\sigma(A_2)$ = Error in $A_2$;
    $\text{d}a/d\text{t}$ = Semi-major axis drift obtained by converting $A_2$;
    $\sigma(\text{d}a/d\text{t})$ = Error in the semi-major axis drift;
    $Y_M$ = 95-th percentile of the maximum expected Yarkovsky drift;
    S/N  = Signal-to-noise ratio of the detection;
    Res. = Acceptance flag, can be only accepted (Acc.) or rejected (Rej.);
    \#OO = Number of Optical Observations;
    \#ReO  = Number of rejected Optical Observation in the 7-D OD;
    \#ReO$_6$    = Number of rejected Optical Observation in the 6-D OD;
    \#RaO = Number of Radar Observations;
    \#ReRa   = Number of rejected Radar Observations in the 7-D OD;
    \#ReRa$_6$    = Number of rejected Radar Observations in the 7-D OD;
    $D_{15}$ = 15-th percentile of the diameter distribution;
    $D_{50}$ = 50-th percentile of the diameter distribution;
    $D_{85}$ = 85-th percentile of the diameter distribution;
    $P$ = Rotation period;
    Tax = Taxonomic class;
    $\Delta T$ = Length of the observational arc.
    The full list of results in extended precision is available at the CDS.
    The $A_2$ and $\sigma(A_2)$ values are given in $10^{-14}$ au d$^{-2}$, while the values of the semi-major axis drifts $\text{d}a/d\text{t}, \sigma(\text{d}a/d\text{t})$ and $Y_M$ are given in $10^{-4}$ au My$^{-1}$. The diameter percentiles $D_{15}, D_{50},$ and $D_{85}$ are given in meters, the rotation period $P$ in hours, and the arc length $\Delta T$ in years. }
    \label{tab:det_short}
\end{table*}

\begin{table*}[hbt!]
    \renewcommand{\arraystretch}{1.3}
    \centering
    \caption{Values of $\varepsilon_1, \varepsilon_2$, and $\chi$ for the asteroids with $\chi \geq 1$. }
    \tiny
    \begin{tabular}{crrrrrrcccc}
\hline
\hline
   Ast.   &     $A_2$  & $\sigma_{A_2}$  &   S/N  &        $A_2^{\text{JPL}}$   & $\sigma_{A_2}^{\text{JPL}}$ &   S/N$^{\text{JPL}}$  & JPL Model &  $\varepsilon_1$  &  $\varepsilon_2$  &  $\chi$   \\
\hline
65717     &   4.33  &  0.99     &    4.3 &    7.74    &    1.04   &    7.4 &       1   &  3.42  &       3.26    &    2.36   \\
2003 AF$_{23}$  &  $-$4.44  &  0.32     &   13.7 &   $-$5.48    &    0.34   &   15.6 &       1   &  3.21  &       2.96    &    2.18   \\
2011 PU$_1$  &  $-$44.44 &  6.68     &    6.6 &   $-$30.30   &    4.99   &    6.0 &       1   &  2.11  &       2.83    &    1.69   \\
2062      &  $-$1.40  &  0.16     &    8.3 &   $-$1.06    &    0.11   &    9.2 &       1   &  2.04  &       3.00    &    1.69   \\
1998 KY$_{26}$  &  $-$22.99 &  3.96     &    5.7 &   $-$13.77   &    5.66   &    2.4 &       3   &  2.32  &       1.62    &    1.33   \\
4179      &  $-$0.61  &  0.06     &    8.8 &   $-$0.77    &    0.09   &    7.9 &       2   &  2.24  &       1.60    &    1.30   \\
2018 CW$_2$   &  $-$49.22 &  2.39     &   20.5 &   $-$55.01   &    3.85   &   14.2 &       1   &  2.41  &       1.50    &    1.27   \\
1685      &  $-$0.44  &  0.10     &    4.3 &   $-$0.30    &    0.04   &    6.5 &       1   &  1.39  &       3.08    &    1.27   \\
163000    &  $-$3.48  &  0.55     &    6.2 &   $-$2.46    &    0.60   &    4.0 &       1   &  1.84  &       1.70    &    1.25   \\
85990     &  $-$3.15  &  0.21     &   14.4 &   $-$3.56    &    0.26   &   13.4 &       1   &  1.87  &       1.54    &    1.19   \\
2018 GY    &  $-$30.67 &  4.18     &    7.3 &   $-$39.49   &    6.32   &    6.2 &       1   &  2.10  &       1.39    &    1.16   \\
306383    &  $-$2.50  &  0.44     &    5.6 &   $-$1.71    &    0.54   &    3.1 &       1   &  1.79  &       1.47    &    1.13   \\
2010 VK$_{139}$ &  $-$14.71 &  2.12     &    6.9 &   $-$9.81    &    3.80   &    2.5 &       1   &  2.30  &       1.28    &    1.12   \\
363599    &  $-$6.37  &  0.49     &   12.8 &   $-$5.51    &    0.60   &    9.1 &       1   &  1.72  &       1.42    &    1.09   \\
2005 VL$_1$   &  $-$21.68 &  2.33     &    9.2 &   $-$83.20   &   56.12   &    1.4 &       3   & 26.36  &       1.09    &    1.09   \\
524522    &  $-$5.98  &  0.10     &   56.9 &   $-$5.82    &    1.12   &   51.8 &       1   &  1.49  &       1.40    &    1.02   \\ 
\hline
    \end{tabular}
        \tablefoot{Columns are:
     Ast = Designation of the asteroid;
   $A_2$ = Transversal nongravitational component estimated by NEOCC;
   $\sigma_{A_2}$ = Error in $A_2$ estimated by NEOCC;
   S/N = Signal-to-noise ratio of the NEOCC detection;
   $A_2^{\text{JPL}}$   = Transversal nongravitational component estimated by JPL;
   $\sigma_{A_2}^{\text{JPL}}$ = Error in $A_2$ estimated by JPL;
   S/N$^{\text{JPL}}$  = Signal-to-noise ratio of the JPL detection;
   JPL Model = Dynamical model used by JPL (1 = $A_2$; 2 = $A_1$ and $A_2$; 3 = $A_1, A_2$ and $A_3$);
   $\varepsilon_1$ = Relative error w.r.t. NEOCC;
   $\varepsilon_2$ = Relative error w.r.t. JPL;
   $\chi$ = Chi value.
   The full list in extended precision is available at the CDS.
    The $A_2$ and $\sigma(A_2)$ values are given in $10^{-14}$ au d$^{-2}$. }
    \label{tab:chivalues}
\end{table*}

\begin{table*}[hbt!]
    \renewcommand{\arraystretch}{1.1}
    \centering
    \caption{Asteroids with measured Yarkovsky effect for the JPL SBDB, not identified or rejected by our procedure.}
    \begin{tabular}{crrrrrrrc}
\hline
\hline
Ast.       &  $\text{d}a/d\text{t}$ &    $\sigma(\text{d}a/d\text{t})$ &   S/N  &  $Y_M$  & $\text{d}a/d\text{t}^{\text{JPL}}$ &    $\sigma(\text{d}a/d\text{t}^{\text{JPL}})$ & S/N$^{\text{JPL}}$    &  JPL Model  \\
\hline
523599    &  829.24       &  19.95              &  41.5   &  21.83     &   1292.37 &       22.09   &  58.4  &       1  \\
66391     &  $-$3.12        & 2.25              &   1.384 &  13.53     &  $-$5.47    &       0.29    &  18.7  &       1  \\
2006 CT   &  $-$38.88     &  2.08            &  18.67  &  94.22   & $-$38.51  &       2.09 &  18.39 &       1  \\
65803     &  16.41         & 6.39         &     2.57      &  12.46         &  $-$4.04    &       0.25    &  16.0  &       1  \\
2011 CL$_{50}$ &  1949.59      &  138.59             &  14.0   &  340.71    &   2023.44 &       135.40  &  14.9  &       1  \\
2009 BD   &  $-$667.46      &  20.97              &  31.8   &  285.39    &  $-$473.66  &       34.58   &  13.6  &       2  \\
2004 JN$_1$  &  $-$164.31      &  11.51              &  14.2   &  123.80    &  $-$159.28  &       11.80   &  13.4  &       1  \\
2001 BB$_{16}$ &  153.56       &  11.37              &  13.4   &  133.76    &   150.48  &       12.25   &  12.2  &       1  \\
613995    &  $-$95.23       &  8.76               &  10.8   &  63.79     &  $-$94.35   &       9.00    &  10.4  &       1  \\
3200      &  $-$6.90   & 2.48   &         2.78     &  4.19         &  $-$12.11   &       1.44    &   8.3  &       1  \\
2006 RH$_{120}$&  $-$54628.00     &  1438.9 &  37.96      &  1406.22         &$-$21205.10  &       2664.47 &   7.9  &       3  \\
2013 GZ$_{7}$  &  $-$24.48     &  2.82            &  8.67 & 103.88    &$-$23.11   &       2.97 &   7.78 &       1  \\
2000 AA$_6$  &  71.40        &  9.46               &  7.5    &  40.78     &   76.07   &       10.00   &   7.6  &       1  \\
87024     & 9.41  &  5.83     &     1.61      &  27.04         &   9.33    &       1.25    &   7.4  &       1  \\
2008 DB   &  $-$588.54      &  65.53              &  8.9    &  222.81    &  $-$455.24  &       62.16   &   7.3  &       1  \\
2012 UU$_{136}$&  40.45      &  6.00           &  6.73   &  60.55   &   40.58  &       6.09 &   6.65 &       1  \\
2018 KC$_2$  &  127.86       &  18.24              &  7.0    &  97.81     &   122.98  &       18.69   &   6.5  &       1  \\
2009 EM$_1$  &  $-$118.36      &  14.56              &  8.1    &  62.99     &  $-$122.27  &       18.62   &   6.5  &       1  \\
162004    &  20.48        &  2.29               &  8.9    &  16.18     &   18.31   &       2.78    &   6.5  &       1  \\
2018 HG$_2$  &  $-$186.27      &  18.76              &  9.9    &  152.22    &  $-$186.85  &       29.04   &   6.4  &       1  \\
2015 XC$_{352}$&  $-$323.04      &  49.74              &  6.4    &  218.19    &  $-$332.22  &       55.54   &   5.9  &       1  \\
337248    &  $-$24.04       &  4.13               &  5.8    &  11.33     &  $-$23.73   &       4.64    &   5.1  &       1  \\
2013 BA$_{74}$ &  307.24       &  0.71               &  5.0    &  145.09    &   619.07  &       124.18  &   4.9  &       1  \\
267759    &  $-$11.43     &  2.31            &  4.94   &  19.04     &  $-$10.99 &       2.31 &   4.74&       1  \\
491007    &  $-$42.92     &  9.04            &  4.74  &  38.79   &  $-$40.38 &       9.16 &   4.40&       1  \\
68950     &  $-$2.02   & 1.04      &    1.93      &  8.97         &  $-$2.99    &       0.68    &   4.3  &       1  \\
397326    &  9.76      &  2.05            & 4.75  &  18.84   &  9.86  &       2.26 &  4.35 &       1  \\
2012 TC$_4$  &  $-$16.48 &   16.32  &  1.01      &  1165.75         &  $-$106.88  &       26.05   &   4.1  &       2  \\
2004 FH   &   50.16   & 31.56       &    1.58    &  374.62         &   85.37   &       21.15   &   4.0  &       1  \\
65679     & $-$18.50      & 4.38             & 4.22 & 17.96    & $-$19.71  &       4.89  &   4.02&       1  \\
153201    &  $-$23.52       &  4.22               &  5.5    &  15.80     &  $-$18.59   &       4.69    &   3.9  &       1  \\
5381      &  $-$0.18   & 2.94         & 0.060 &  12.23         &   2.70    &       0.69    &   3.9  &       1  \\
7482      &  1.35  &  3.96   &       0.34     &  8.15         &  $-$2.22    &       0.61    &   3.6  &       1  \\
526826    &  $-$56.99  &  19.35          & 2.94     &  74.76         &  $-$66.69   &       18.68   &   3.5  &       1  \\
163899    &  $-$1.42   & 2.51  &  0.56      &  4.93         &  $-$6.94    &       2.01    &   3.4  &       1  \\
5604      & $-$12.27      & 3.90          & 3.14 &  16.03   &  $-$14.00 &       4.20 &   3.33&       1  \\
2003 MS$_2$  &  $-$22.11  & 7.50        & 2.94      &  61.85         &  $-$27.63   &       8.36    &   3.3  &       1  \\
2021 QH$_2$  &  407.83       &  95.41              &  4.2    &  219.79    &   354.44  &       107.61  &   3.2  &       1  \\
2016 GE$_1$  &  $-$394.54      &  112.11             &  3.5    &  123.95    &  $-$582.52  &       177.24  &   3.2  &       1  \\
25143     &  $-$7.84 & 4.86         & 1.61     &  24.69         &  $-$11.09   &       3.38    &   3.2  &       1  \\
2014 QL$_{390}$&  $-$45.92  & 16.27          & 2.82     &  52.65         &  $-$51.76   &       15.99   &   3.2  &       1  \\
138947    &  $-$13.20   &  4.61          & 2.86      &  21.90         &  $-$18.10   &       5.67    &   3.1  &       1  \\
369984    &  $-$52.58   &26.60          & 1.97     &  23.11         &  $-$11.65   &       3.68    &   3.1  &       1  \\
22753     &  $-$8.69 & 3.345         & 2.59     &  19.28         &  $-$10.35   &       3.28    &   3.1  &       1  \\
483423    &  $-$51.54  &  18.22          & 2.82      &  88.76         &  $-$58.96   &       18.97   &   3.1  &       1  \\
396593    &  12.29    & 4.56         & 2.69      &  29.31         &   14.89   &       4.83    &   3.0  &       1  \\
417210    &   $-$15.02  & 5.23         & 2.87      &  54.46         &  $-$16.46   &       5.34    &   3.0  &       1  \\
22099     &  3.19  &  2.27        &  1.40     &  16.45         &   5.02    &       1.65    &   3.0  &       1  \\
\hline
    \end{tabular}
        \tablefoot{Columns are:
         Ast = Designation of the asteroid;
  $\text{d}a/d\text{t}$ = Semi-major axis drift computed by the NEOCC; 
  $\sigma(\text{d}a/d\text{t})$ = Uncertainty in the semi-major axis drift computed by the NEOCC;
  S/N  = Signal-to-noise ratio of semi-major axis drift detection of NEOCC measurement;
  $Y_M$ = 95-th percentile of the expected maximum Yarkovsky semi-major axis drift;
  $\text{d}a/d\text{t}^{\text{JPL}}$ = Semi-major axis drift computed by JPL;
  $\sigma(\text{d}a/d\text{t}^{\text{JPL}})$ = Uncertainty in the semi-major axis drift computed by JPL;
  S/N$^{\text{JPL}}$    = Signal-to-noise ratio of semi-major axis drift detection of JPL measurement;
  JPL Model = Dynamical model used by JPL (1 = $A_2$; 2 = $A_1$ and $A_2$; 3 = $A_1, A_2$ and $A_3$).
   The full list in extended precision is available at the CDS.
        The values of $\text{d}a/d\text{t}, \sigma(\text{d}a/d\text{t})$ and  $Y_M$  are given in $10^{-4}$ au My$^{-1}$. }
    \label{tab:jpl_comparison}
\end{table*}

\begin{table*}[hbt!]
    \renewcommand{\arraystretch}{1.05}
    \centering
    \caption{Asteroids with measured Yarkovsky drift with isolated tracklets at the beginning of the observational arc.}
    \small
    \begin{tabular}{cccrrrrrrrc}
    \hline
    \hline
         Ast.              &    Tracklet Initial Date        &  Obs. Code & Arc length &   $\Delta t$    &    $A_2$    &  $\sigma(A_2)$    &    $\text{d}a/\text{d}t$       &   $\sigma(\text{d}a/\text{d}t)$ &  S/N    & Except \\
\hline
2007~TF$_{15}$    &    2007-10-09  &  G96 &  1  &    6.9  & $-$22.55 &  2.11   &   $-$91.13    &    8.54 & 10.67   & N \\
326290            &    1993-04-13  &  691 & <1  &    5.0  &  $-$3.17 &  0.32   &   $-$17.85    &    1.83 &  9.72   & N \\
4179              &    1934-02-14  &  012 &  4  &   42.2  &  $-$0.61 &  0.06   &    $-$2.72    &    0.30 &  8.81   & N \\
175706            &    1985-12-18  &  675 &  1  &   10.2  &  $-$1.20 &  0.14   &    $-$5.69    &    0.67 &  8.47   & N \\
2062              &    1955-12-17  &  675 & <1  &   20.0  &  $-$1.43 &  0.17   &    $-$6.42    &    0.77 &  8.27   & N \\
481442            &    2001-11-12  &  645 & <1  &    5.0  &  $-$5.64 &  0.70   &   $-$33.50    &    4.19 &  7.98   & N \\
385186            &    1986-12-29  &  381 & <1  &    7.0  &     1.18 &  0.14   &       4.80    &    6.05 &  7.93   & N \\
350751            &    1991-03-09  &  261 & <1  &   10.7  &  $-$5.20 &  0.65   &   $-$22.87    &    2.89 &  7.91   & N \\
85953             &    1971-03-26  &  675 & <1  &   27.9  &  $-$1.51 &  0.19   &   $-$14.82    &    1.87 &  7.89   & N \\
152563            &    1953-01-12  &  261 &  2  &   39.0  &  $-$2.96 &  0.39   &   $-$14.25    &    1.92 &  7.42   & Y \\
66400             &    1987-04-21  &  675 & <1  &   12.1  &  $-$2.33 &  0.36   &   $-$15.96    &    2.47 &  6.44   & N \\
1994~GL           &    1994-04-10  &  691 &  3  &   25.9  &    45.31 &  7.16   &      11.96    &    9.35 &  6.32   & N \\
162361            &    1991-01-15  &  675 & <1  &    8.9  &     2.89 &  0.47   &      15.80    &    2.60 &  6.06   & N \\
2019~QU$_2$       &    2004-08-12  &  807 & <1  &    6.0  &  $-$6.80 &  1.25   &   $-$63.26    &   11.68 &  5.41   & N \\
446924            &    1992-10-02  &  261 & <1  &   10.0  &  $-$3.37 &  0.72   &   $-$19.32    &   4.15 &  4.65   & N \\
138404            &    1982-07-20  &  260 & <1  &    6.6  &     2.09 &  0.46   &       9.26    &   2.05 &  4.50   & N \\
234341            &    1991-04-18  &  260 & <1  &    8.8  &  $-$3.28 &  0.72   &   $-$22.59    &   5.01 &  4.50   & N \\
511684            &    2005-12-03  &  568 & <1  &    9.1  &    10.02 &  2.34   &      62.68    &  14.63 &  4.28   & N \\
283457            &    1951-11-05  &  261 & <1  &   39.7  &  $-$7.05 &  1.66   &   $-$25.34    &   5.99 &  4.23   & N \\
203471            &    1994-01-08  &  675 &  1  &    8.0  &  $-$7.77 &  1.86   &   $-$41.49    &   9.97 &  4.16   & N \\
85774             &    1989-04-30  &  260 & <1  &    9.4  &  $-$0.53 &  0.12   &    $-$2.15    &   0.52 &  4.12   & N \\
7350              &    1963-10-15  &  675 & <1  &   22.5  &  $-$1.86 &  0.49   &    $-$8.03    &   2.12 &  3.77   & N \\
592968            &    2003-03-24  &  644 & <1  &   12.0  &     9.28 &  2.47   &      64.55    &  17.23 &  3.74   & N \\
85770             &    1990-10-12  &  261 & <1  &    8.0  &  $-$5.03 &  1.36   &   $-$24.26    &   6.60 &  3.67   & N \\
162181            &    1979-12-20  &  413 & <1  &   19.4  &  $-$1.90 &  0.57   &    $-$7.40    &   2.24 &  3.29   & N \\
398188            &    2001-07-02  &  644 & <1  &    8.9  &  $-$3.01 &  0.91   &   $-$14.87    &   4.51 &  3.29   & N \\
2017~DN$_{109}$   &    2010-05-02  &  C51 & <1  &    6.7  &    11.87 &  3.66   &     145.86    &  45.02 &  3.23   & N \\
7341              &    1981-10-23  &  095 & <1  &    9.8  &  $-$0.56 &  0.17   &    $-$2.38    &   0.73 &  3.23   & N \\
138911            &    1984-08-02  &  260 & <1  &    8.1  &  $-$2.49 &  0.80   &    $-$9.19    &   2.96 &  3.10   & N \\
162080            &    1981-03-03  &  413 &  1  &   16.9  &  $-$2.28 &  0.74   &   $-$11.76    &   3.81 &  3.08   & N \\
474158            &    1978-10-10  &  260 & <1  &   19.8  & $-$10.39 &  3.50   &   $-$43.28    &  14.60 &  2.96   & N \\
65679             &    1954-11-20  &  675 & <1  &   34.9  &  $-$3.10 &  1.04   &   $-$14.84    &   5.00 &  2.96   & N \\
2006CT            &    1991-01-15  &  691 & <1  &   13.6  &  $-$7.58 &  2.56   &   $-$32.48    &  10.99 &  2.95   & N \\
469219            &    2004-03-17  &  645 & <1  &    7.0  & $-$11.73 &  4.50   &   $-$50.33    &  19.31 &  2.60   & N \\
2012~BB$_{124}$   &    1976-08-27  &  260 & <1  &   35.4  &     6.64 & 2.909   &      27.48    &  12.03 &  2.28   & Y \\
2012~GA$_5$       &    2001-03-20  &  645 & <1  &   10.9  &    42.33 & 19.68   &     168.96    &  78.59 &  2.14   & N \\
491007            &    2000-04-28  &  699 & <1  &   10.9  & $-$11.77 &  5.89   &   $-$93.69    &  46.90 &  1.99   & N \\
267759            &    1953-04-13  &  261 & <1  &   44.7  &  $-$3.03 &  2.01   &   $-$11.37    &   7.55 &  1.50   & N \\
26310             &    1951-07-08  &  261 & <1  &   47.2  & $-$11.79 &  7.99   &   $-$43.89    &  29.75 &  1.47   & N \\
358453            &    1994-03-06  &  691 & <1  &   13.0  &  $-$6.09 &  4.91   &   $-$30.33    &  24.44 &  1.24   & N \\
510190            &    2002-01-09  &  645 & <1  &    9.0  &    7.242 &  6.45   &      30.12    &  26.86 &  1.12   & N \\
5604              &    1976-04-23  &  413 & <1  &    8.9  &  $-$0.85 &  0.84   &    $-$4.49    &   4.44 &  1.01   & N \\
2018~KR           &    2013-04-19  &  F51 & <1  &    5.0  & $-$53.82 & 52.94   &  $-$254.06    & 249.88 &  1.01   & Y \\
2013~GZ$_7$       &    2003-03-31  &  645 & <1  &   10.0  &  $-$2.96 &  3.43   &   $-$11.03    &  12.77 &  0.86   & N \\
2015~XW$_{261}$   &    2009-01-16  &  645 & <1  &    6.8  &    32.78 & 39.13   &     159.35    & 190.21 &  0.83   & N \\
2014~RQ$_{17}$    &    1953-10-08  &  261 & <1  &   60.9  &    65.74 & 80.30   &     261.30    & 319.14 &  0.81   & Y \\
403775            &    2002-06-02  &  644 &  1  &    7.9  &  $-$2.77 &  4.37   &   $-$22.61    &  35.67 &  0.63   & N \\
2012~DY$_{32}$    &    2003-03-22  &  644 & <1  &    8.1  &    37.01 & 76.56   &     155.24    & 321.12 &  0.48   & N \\
339492            &    1983-10-26  &  260 & <1  &   16.5  &     1.83 & 3.943   &       6.85    &  14.70 &  0.46   & N \\
2021~SP$_4$       &    2005-09-28  &  645 & <1  &    6.9  &  $-$8.47 &  24.1   &   $-$35.54    & 101.45 &  0.35   & N \\
2012~UU$_{136}$   &    2002-11-08  &  645 & <1  &    9.9  &  $-$2.47 &  7.10   &   $-$10.20    &  29.32 &  0.34   & N \\
2015~LA$_2$       &    2010-05-16  &  F51 &  5  &    5.0  &    54.38 &171.79   &     266.77    & 842.74 &  0.31   & Y \\
2016~GL$_2$       &    2009-04-02  &  G96 &  1  &    7.0  &    17.64 & 94.08   &      66.90    & 356.78 &  0.18   & N \\
397326            &    1953-03-09  &  675 & <1  &   53.5  &     9.77 &  5.27   &       3.68    &  19.87 &  0.18   & N \\
375505            &    1998-11-17  &  645 & <1  &    9.9  &  $-$4.98 &  2.69   &    $-$1.81    &   9.80 &  0.18   & N \\
2016~UW$_{80}$    &    2011-10-25  &  F51 & <1  &    5.0  &  $-$6.43 &579.31   &  $-$260.58    &2347.70   &  0.11   & Y \\
465617            &    2001-12-18  &  645 & <1  &    7.2  &  $-$6.65 &  6.43   &    $-$2.67    &  25.89 &  0.10   & N \\
2005~TF$_{49}$    &    1989-10-21  &  260 & <1  &   15.9  &     2.67 &  3.77   &       1.11    &  15.70 &  0.07   & N \\
\hline
    \end{tabular}
        \tablefoot{Columns are: Ast. =  Designation of the asteroid;
Tracklet Initial Date = Date of the first observation in the tracklet;
Obs. Code = Observatory code;
Arc length: Arc length of the isolate tracklet;
$\Delta t$: Time passed between the isolated tracklet and the following available observation;
$A_2$ = Value of $A_2$ obtained by excluding the isolated tracklet;
$\sigma(A_2)$ = Error in $A_2$ obtained by excluding the isolated tracklets;
$\text{d}a/\text{d}t$ = Value of $A_2$ obtained by excluding the isolated tracklet;
$\sigma(\text{d}a/\text{d}t)$ = Error in $A_2$ obtained by excluding the isolated tracklets;
S/N = Signal-to-noise obtained by excluding the isolated tracklets;
Except = Flag to indicate that the tracklet is reliable. The arc length is given in days, the time $\Delta t$ in years, the values of $A_2, \sigma(A_2)$ in $10^{-14}$ au d$^{-2}$, and $\text{d}a/\text{d}t, \sigma(\text{d}a/\text{d}t)$ in $10^{-4}$ au My$^{-1}$. }
    \label{tab:validation_final}
\end{table*}

\end{appendix}

\end{document}